\documentstyle[aps,prb]{revtex} 
\input epsf.sty 
\renewcommand{\theequation}{\arabic{section}.\arabic{equation}} 
 
\begin{document} 
 
\twocolumn[ 
\hsize\textwidth\columnwidth\hsize\csname@twocolumnfalse\endcsname 
 
\draft 
 
\title{Effective Gauge Field Theory of  the 
 t-J Model in the\\   Charge-Spin Separated State and    
its   Transport Properties} 
\author{Ikuo Ichinose\cite{ichinose}} 
\address{Institute of Physics, University of Tokyo, 
Komaba, Tokyo, 153-8902 Japan} 
\author{Tetsuo Matsui\cite{matsui}} 
\address{Department of Physics, Kinki University, 
Higashi-Osaka, 577-8502 Japan 
} 
\author{Masaru Onoda\cite{onoda}} 
\address{High Energy Accelerator Research Organization (KEK),
Tsukuba, 305-0801 Japan
} 
\date{\today}  
 
\maketitle 
 
\begin{abstract}   
We study the slave-boson t-J model of cuprates with high 
superconducting transition temperatures, and derive its  
low-energy effective field theory for the charge-spin  
separated state in a self-consistent manner.  
The phase degrees of freedom of the mean field for hoppings  
of holons and spinons can be regarded as a U(1) gauge field, 
$A_i$. The charge-spin separation occurs below certain  
temperature, $T_{\rm CSS}$, as a deconfinement phenomenon 
of       the dynamics of $A_i$. 
Below certain temperature $T_{\rm SG} ( < T_{\rm CSS})$,  
the spin-gap phase develops as the Higgs phase of the  
gauge-field dynamics, and $A_i$ acquires a mass $m_A$. 
The effective field theory  near $T_{\rm SG}$
takes the form of Ginzburg-Landau 
theory of a complex scalar field $\lambda$ coupled with $A_i$, 
where $\lambda$ represents $d$-wave pairings of spinons.   
Three dimensionality of the system is crucial to 
realize a phase transition at $T_{\rm SG}$.
 By using this field theory, we calculate  
the dc resistivity $\rho$. At $T > T_{\rm SG}$, $\rho$ is
proportional to $T$. At  $T < T_{\rm SG}$, it  deviates  
downward from the $T$-linear behavior    
as $\rho \propto T \{ 1 -c(T_{\rm SG}-T)^d \}$. 
When the system is near (but not) two dimensional,  
due to the compactness of the phase of the field $\lambda$, 
the exponent $d$ deviates from its mean-field value $1/2$ and 
becomes a nonuniversal quantity which depends on temperature  
and doping. This significantly improves the comparison with  
the experimental data. 
\end{abstract} 
 
\pacs{74.25.Fy, 71.27.+a, 71.10.Pm, 11.15.-q} 
 
]
 
\setcounter{footnote}{0} 
\section{Introduction} 
 
In many physical systems,  one may identify their microscopic  
Hamiltonians, but calculations of physical quantities starting  
from these Hamiltonians may not be straightforward. 
One promising way to overcome this difficulty is to derive  
an effective theory that is appropriate for the  
energy scale in question and use it for calculations. 
 
The discovery of cuprates with high superconducting transition  
temperatures $T_c$ has stimulated condensed-matter  
physicists for one and a half decades. 
Here  the t-J model serves as a canonical model of cuprates, and 
is expected to have potentiality to explain their various 
interesting properties observed in experiments. 
The t-J model  itself can be regarded as a low-energy effective  
model of the  Hubbard model or the d-p model at the energy scales 
of  the hopping amplitude of electrons  $t (\simeq 0.3 $eV),    
and  the antiferromagnetic (AF) spin couplings $J (\simeq 0.1$eV),  
which are much lower than the energy scales of 
the strong onsite Coulomb repulsion $U (\simeq 3$eV).  
The t-J model allows us to understand some basic physical  
properties of strongly-correlated electrons. For example,   
the origin of the superconductivity is attributed to the formation
and condensation of hole  pairs at nearest-neighbor (NN) sites 
in the background of short-range AF spin order.\cite{IM}  
Furthermore,  the  mean-field theory (MFT) in the slave-boson (SB)  
representation predicts an interesting  phase diagram in the doping  
($\delta$)-temperature ($T$) plane such as the spin-gap state in  
which the NN AF spin pairings develop.  
However, it is still not easy to calculate physical 
quantities such as dc resistivity analytically  
from the t-J model itself. Thus it is welcome to derive an effective  
theory of the t-J model at lower energies. It is preferable that 
this effective theory takes a form of local field theory so that many 
established techniques are applicable.  
 
Many quasi-two-dimensional cuprates  exhibit certain anomalous 
metallic behavior above $T_c$ in various quantities such as dc 
resistivity, Hall coefficient, magnetic susceptibility.\cite{sato}
For example, the dc resistivity $\rho$ at  fixed $\delta$ shows 
the $T$-linear dependence \cite{fiory} on $T$ above certain temperature.  
These anomalies call for a new theoretical explanation, probably in   
a framework beyond the conventional Fermi-liquid theory. 
Anderson \cite{anderson} pointed out  that the charge-spin separation 
(CSS) phenomenon may explain them.  
 
The CSS is a phenomenon in which charge and spin degrees of freedom of  
strongly-correlated electrons  behave independently. 
In the SB representation, an electron is viewed as a composite 
of two constituents, a holon and a spinon, which bear  
charge and spin degrees of freedom of electron respectively. 
The CSS is ``naturally" described by the SB   
(or slave-fermion) MFT, in which the holons and spinons       
are treated  as quasifree particles having no correlations among them. 
When one incorporates fluctuations of phase degrees of freedom of MFs,   
these phases behave as  gauge fields  coupled to  holons and spinons, 
and the system possesses a local U(1) gauge symmetry.  

In the previous papers\cite{imcss} (hereafter we  call them Papers I),  
we have studied the possibility of the CSS in the t-J model using  
gauge-theoretical methods. \cite{gauge}
By introducing auxiliary fields, the system
can be viewed as a lattice gauge model.
The corresponding gauge dynamics  
has two general possibilities in its realization; 
(i) the confining phase in which holons and spinons are confined   
to electrons and the gauge fields fluctuate strongly, or (ii) the  
deconfining phase in which holons and spinons   
appear as almost independent quasi-free particles and the 
gauge-field fluctuations 
are weak and can be treated by the usual perturbation theory. 
We interpreted the CSS as the second possibility above; 
CSS is a deconfinement phenomenon of gauge theory.  
Furthermore, we obtained the result that the system exhibits a  
confinement-deconfinement phase transition (CDPT) at certain 
transition temperature $T_{\rm CSS}$, 
below which the deconfinement phase takes place, 
i.e., the CSS takes place below  $T_{\rm CSS}$.  
The CDPT is of second order when the system has  three-dimensional 
couplings   whatever small they are, while it is infinite order 
of Kosterlitz-Thouless type when the system is purely two dimensional. 
  
Very recently, study of the  CSS was revived, 
and the confinement-deconfinement problem of gauge theories 
of nonrelativistic fermions is addressed.
Unfortunately, most of these studies do not quote the previous works 
in which important results on that problem were obtained. 
Moreover, some of the recent arguments are in apparent 
contradictions to the previous results, but these authors do not 
discuss the origins of these contradictions. 
For example, in his recent paper\cite{nayak} Nayak discussed 
the above problem and concluded that the slave particles (holons
and spinons) are always confined by  U(1) gauge field. 
This result is obviously in sharp contrast to our result in
 Papers I which predicts a deconfinement phase at low $T$. 
 
How does this contradiction come out? 
In Nayak's paper,  dynamics of the gauge field is {\em not} 
discussed at all and it is simply concluded that the infinitely 
strong gauge coupling in the bare Lagrangian necessarily leads 
to confinement.
On the other hand, in Papers I, we studied the gauge-field dynamics 
by using nonperurbative methods. Even if the gauge coupling is 
infinite in the original lattice model, 
the gauge field can acquire nontrivial dynamics at low energies
such as a deconfinement phenomenon due to the couplings to matter 
fields. In short,  due to ample pair creations of  holons and/or 
spinons, fluxes of electric-like field connecting external charges 
which would cause the 
linear-rising confining potential are truncated into short segments,  
giving rise to a deconfining potential. 
In lattice gauge theory such as multi-flavor QCD,\cite{iwasaki}
this fact is now well-established and verified by numerical studies.  
We showed that a similar deconfinement phenomenon occurs in the 
gauge theory of the slave-particle t-J model.  

Above is the main point of our comment \cite{nayakcomment}
to Nayak's paper.\cite{nayak} 
Another point we argued is that the
Elitzur's theorem \cite{Elitzur} does not exclude the possibility of
deconfinement phase.  
There are some misunderstandings 
that the Elitzur's theorem prohibits the existence of the 
deconfinement phase, since the theorem states that the averages 
of gauge-noninvariant quantities should vanish. However, 
MFT, for example, can describe the deconfinement phase in accord with 
the Elizur's theorem by averaging over the gauge-rotated
copies of a MF solution.\cite{Drouffe}

In his reply,\cite{nayakreply} 
Nayak admitted that these two points are certainly correct. 
There he also posed an argument \cite{nayakreply} that
there are ambiguities in assigning EM charges to slave particles,
which makes the possibility of CSS quite doubtful.
This is a misunderstanding. Our explanation is as follows;  
This ambiguity is related to the choice of reference state 
from which the EM charges are
measured. If measured from the vacuum, the EM charge of a 
holon is zero and the charge of a spinon is the 
same as the charge of an electron, $e ( < 0)$. 
Similarly, if they are measured from the half-filled state,
the holon charge is $-e$ and the spinon charge is zero.
Since this is an important
point to understand the slave particle approach, we present
our explanation in Sect.6 of the present paper \cite{ourexp}
after deriving the relevant formulae in Sect.5A.

In the present paper, we focus on the CSS state in the SB t-J model 
and derive the low-energy effective field theory.  
By using the lattice model suggested by this field theory, we
confirm that the nature of the transition into the spin-gap state is
not a crossover but a genuine phase transition for the 
three-dimensional system. We apply this effective field theory to 
calculate the dc resistivity. 
Previously, Lee and Nagaosa \cite{ln} and Ioffe et al. \cite{iw}
showed that $\rho \propto T$ 
for fermions and bosons interacting with  a massless gauge field.  
Their  system has some relation to the  uniform RVB MFT 
of the SB t-J model.  
Recent experiments on YBaCuO by Ito {\it et al.}\cite{uchida} 
and others reported that $\rho$ deviates downward from the 
$T$-linear behavior below certain temperature $T_{\rm SG} ( > T_c)$. 
This  $T_{\rm SG}$ coincides with the temperature determined by 
NMR and neutron experiments at which a spin gap starts to 
develop.\cite{sgexp}   
So it is quite interesting to derive an effective theory  
in the spin-gap state of  the SB t-J model, and calculate the 
spin-gap effect on $\rho$ as  a function of $\delta$ and $T$.  
The effective theory used in Ref.\cite{ln} is inadequate 
for this purpose, since it assumes no spin gap.  
In Sect.3, we shall see that the spin gap generates a mass 
of gauge field via the Anderson-Higgs mechanism. Thus the effect 
of scatterings of holons and spinons by gauge fields is reduced. 
This gives rise to a downward deviation of $\rho$ from the 
$T$-linear behavior. Our explicit result of $\rho$ 
in Sect.5C is consistent with the experiments.\cite{uchida}  

The strucutre of the present paper is as follows;
In Sect.2, we start with the MFT of the SB t-J model, and review the 
results of Papers I for the U(1) gauge dynamics of the t-J model in a 
compact but selfcontained manner. 
Various analytical and numerical results which are relevant for 
the present system are consulted in order to obtain the correct  
phase diagram.  
When  $T$ is higher than the CDPT temerature 
$T_{\rm CSS}(\delta)$ which depends on the holon density $\delta$,
$T > T_{\rm CSS}(\delta)$, the gauge dynamics is realized in the 
confinemnet phase and the holons and spinons are confined to electrons.
When $T < T_{\rm CSS}(\delta)$, the gauge dynamics is realized 
in the deconfinement phase and the CSS takes place as mentioned.   
The MFT predicts another phase transition at $T_{\rm SG} 
( < T_{\rm CSS} )$, below which the spin gap develops. 
This transition is sometimes claimed to be a crossover rather than
a genuine phase transition.\cite{NL} We shall discuss this problem
in Sect.4 in detail, and show that it is a genuine phase transition
when the system has weak but finite three dimensionality.

In Sect.3, we shall derive the low-energy and low-temperature
effective field theory in the CSS state of the SB t-J model.
The result of Sect.2 assures that this derivation is self consistent,  
since the gauge dynamics is deconfining  at $T < T_{\rm CSS}$. 
The effective theory is a Ginzburg-Landau (GL) theory $L(\lambda, A_i)$  
coupled with a gauge field $A_i$, where the complex scalar $\lambda$  
represents the $d$-wave spinon pairing and $A_i$ is the phase of the
hopping amplitude of holons and spinons. 
Halperin, Lubensky, and Ma \cite{Halperin&Lubensky&Ma} 
considered a similar system, where 
$A_i$  corresponds to the electromagnetic (EM) 
field and $\lambda$ to the Cooper-pair field. 
They calculated the effect of $A_i$ on $\lambda$ to  
conclude that it converts the second-order phase transition to a 
first-order one.  Then it  was pointed out that the radial  
fluctuations of $\lambda$ may keep the transition second-order  
in some parameter region.\cite{Kleinert} This is supported by 
numerical simulations. 
More recently, Ubbens and Lee \cite{Ubbens&Lee} calculated the 
one-loop effect of $A_i$ in the SB MFT of the t-J model, and concluded 
again that the pairing transition at $T_{\rm SG}$  becomes first order. 
(However, their $T_{\rm SG}$ appears below the superconducting 
transition temperature $T_{\rm c}$, so they concluded that the 
spin-gap phase is 
completely destroyed by gauge-field fluctuations.)  
In the present study, we take the compactness of  
the phase of $\lambda$ into account. 
Even in the CSS state, the compactness generates nontrivial 
interaction vertices that are missing in 
the previous treatments.\cite{Halperin&Lubensky&Ma,Ubbens&Lee} 
We find that the periodic interaction above stabilizes the  
system at $T_{\rm SG}$, which is higher than $T_{\rm c}$ at low 
$\delta$.  This allows us to use this effective theory to study 
the spin-gap state at $ T_c < T < T_{\rm SG}$.
We find that the critical exponents  
(such as $d$ that characterizes the gauge field mass $m_A$ through
$m_A \propto (T_{\rm SG} - T)^d$) become $T$ and $\delta$-dependent, 
that is, they are not universal anymore. 

In Sect.4, we study the phase transition into the spin-gap state 
for the 3D system in
detail by using the lattice Abelian Higgs model which is a natural
extension of the effective field theory of Sect.3 to a lattice model.
Below certain on-set temperature $T_{\rm SG}$  
(which is shown to be lower than  $T_{\rm CSS})$, the spin-gap phase 
develops as a Higgs phase of the gauge-field dynamics, 
in which the gauge field acquires a finite mass $m_A$. 
Recent numerical studies on 3D U(1) gauge Higgs model\cite{Higgs} 
show a genuine phase transition and existence of the Higgs phase 
which supports our discussion on the spin-gap phase. 
In the region $ T_{\rm SG} < T < T_{\rm CSS}$, the gauge dynamics is 
realized in the so-called Coulomb phase in which $m_A = 0$. 
At the end of Sect.4, we also comment on the recent dual vortex theory
of bosonized fermions in pure two dimensions
by Balents et al.\cite{BFN}. 
 
In Sect.5, we calculate the dc resistivity $\rho$ as 
an example of calculations of physical quantities in the spin-gap  
state by using the effective field theory obtained in Sect.3.  
First, we obtain the expression of $\rho$ 
in the random-phase approximation (RPA), which depends  
crucially on the $T$-dependence  of $m_A$. 
Then we substitute $m_A$ calculated in Sect.3 to this
$\rho$. 
 
The results on these problems of Sect.3-5  
have been reported in part briefly in our previous two 
short letters.\cite{JPSJ}  
In Sect.3-5 of the present paper, we account for the details of 
these results in a self-contained manner.  

Sect.6 is devoted to discussion. We present a couple of comments 
related to the problem of observing holons and spinons 
in the CSS state and ambiguity of charge assignement for
holon and spinon. These comments and remarks should clarify certain
misunderstandings\cite{nayakreply} on the confinement-deconfinement 
problem in the slave-particle studies of strongly-correlated 
electron systems.

In Appendix, we  consider the 
U(1) Higgs model in the 3D continuum that is tightly related to 
the lattice model we studied in Sect.4. 
We study its phase structure in terms of dual variables, and 
critisize the recent result of Nagaosa and Lee\cite{NLdual}
on this model.

\section{Phase Structure of the Gauge Dynamics} 
\setcounter{equation}{0} 
 
In Sect.2A, we review the MFT of the SB t-J model and 
show the phase diagram in the $\delta$-$T$ plane.
In Sect.2B, we study the effect of fluctuations of MFs in a 
general viewpoint of gauge theory. 
A CDPT takes place
at $T_{\rm CSS}$, below which the CSS occurs.
In Sect.2C, we  briefly explain the spin gap state as a ``dual" 
Higgs mechanism of lattice gauge theory with multiple gauge fields.

\subsection{MFT of the SB t-J Model} 
 
Let us start with the t-J model \cite{tJ} Hamiltonian, 
\begin{eqnarray} 
H & = & \sum_{x,i}\Bigl[-t 
\Bigl(C^{\dagger}_{x+ i\: \sigma}C_{x\sigma}  
+ {\rm H.c.}\Bigr)
\nonumber\\        
 &+& J \Bigl(\vec{S}_{x+i}\cdot\vec{S}_{x}-\frac{1}{4}
 n_{x+i}\ n_x \Bigr)\Bigr],\nonumber\\ 
 \vec{S}_{x} & = & \frac{1}{2}\sum_{\sigma,\sigma'} 
 C^{\dagger}_{x\sigma}\vec{\sigma}_{\sigma \sigma'} 
 C_{x\sigma'},\ \ \ n_x = \sum_{\sigma}C^{\dagger}_{x\sigma}
 C_{x\sigma}, 
\label{htj} 
\end{eqnarray} 
where $C_{x\sigma}$ is the electron 
annihilation operator at the site $x$  
of a  2D or 3D lattice with 
the $z$-component of spin $\sigma  (= \uparrow, \downarrow)$. 
We use $i (=1,2 (,3))$ both as the direction index 
and as the unit vector in the $i$-th direction.\cite{3dcoupling} 
The SB representation of  $C_{x\sigma}$ is written as  
\begin{eqnarray} 
C_{x\sigma} = b_x^{\dagger} f_{x\sigma}, 
\label{slaveboson} 
\end{eqnarray} 
where $b_x$ is the bosonic annihilation operator of holon, and 
$f_{x\sigma}$ is the fermionic annihilation operator of spinon. 
They satisfy the local constraint 
\begin{eqnarray} 
b_x^{\dagger} b_x +\sum_{\sigma}  
f^{\dagger}_{x\sigma}f_{x\sigma} =1, 
\label{constraint} 
\end{eqnarray} 
so that only the three states for each $x$, 
$|{\rm vac}\rangle,$\
$C^{\dagger}_{x\uparrow}|{\rm vac}\rangle,$\
$C^{\dagger}_{x\downarrow}|{\rm vac}\rangle,$\ 
$(C_{x\sigma}|{\rm vac}\rangle = 0)$ are allowed
as physical states and the double-occupancy state 
$C^{\dagger}_{x\uparrow}C^{\dagger}_{x\downarrow}|{\rm vac}\rangle$ 
is excluded
due to the large on-site Coulomb energy $U$.
These three states are 
expressed in the SB representation as 
$b^{\dagger}_x|0\rangle,$\
$f^{\dagger}_{x\uparrow}|0\rangle,$\
$f^{\dagger}_{x\downarrow}|0\rangle,$\ 
$(b_x|0\rangle = f_{x\sigma}|0\rangle = 0)$, respectively. 
The Hamiltonian (\ref{htj})  is rewritten for the physical states
in terms of $b_{x}, f_{x\sigma}$ as 
\begin{eqnarray} 
H&=&-t\sum_{x, i,\sigma} \Bigl(b^{\dagger}_{x+ i}f^{\dagger}_{x \sigma} 
f_{x+ i\:\sigma}b_x+\mbox{H.c.}\Bigr)\nonumber\\ 
&&-\frac{J}{2} \sum_{x, i} \Bigl| 
f^{\dagger}_{x\uparrow}f^{\dagger}_{x+ i\downarrow} 
- f^{\dagger}_{x\downarrow}f^{\dagger}_{x+ i\uparrow}\Bigr|^2. 
\label{Hsb} 
\end{eqnarray} 
This $H$ preserves the total number of holons and spinons at
each $x$, the L.H.S. of (\ref{constraint}), and so maps a physical 
state to another physical state.
In path integral formalism, the partition function 
$Z(\beta) = {\rm Tr} \exp(-\beta H)\; 
[\beta \equiv(k_{\rm B} T)^{-1}]$ is given by 
\begin{eqnarray} 
Z &=& \int [db][df][d\alpha] \:  \exp (-S), \nonumber\\ 
S &=& \int_0^{\beta}d\tau \Bigl[\sum_x \Bigl(  b^{\dagger}_x
\frac{\partial b_x}{\partial\tau}  + 
\sum_\sigma f^{\dagger}_{x\sigma}\frac{\partial f_{x\sigma}}
{\partial\tau} \Bigr) + H + H_{\rm LC}\Bigr], 
\nonumber\\
H_{\rm LC} & = & i\sum_x \alpha_x ( b^{\dagger}_x  b_x +\sum_{\sigma}  
f^{\dagger}_{x\sigma}f_{x\sigma} -1),
\label{ActionS} 
\end{eqnarray} 
where $\tau \ (0 \leq \tau \leq \beta)$ is the imaginary time and
$[db] \equiv \prod_{\tau}\prod_{x}db^{\dagger}_x(\tau)db_x(\tau)$, etc.
Here $b_x(\tau)$ (we omitted the argument $\tau$ in (\ref{ActionS})) 
is regarded as  a complex variable at $x$ and $\tau$,  
and $f_{x\sigma}(\tau)$
is an anticommuting Grassmann variable. 
$H_{\rm LC}$ respects the constraint (\ref{constraint}) via the 
integration over the Lagrange multiplier 
$\alpha_x(\tau)$.\cite{IMconstraint}

To set up the MFT, we introduce two complex auxiliary fields 
$\chi_{xi}$ and $\lambda_{xi}$ on the link $(x, x+i)$ to 
decouple both $t$ and $J$ terms \cite{IMS2} as 
\begin{eqnarray} 
H_{\rm MF}&=&\sum_{x, i}\Bigl[ 
 \frac{3J}{8}\:|\chi_{xi}|^2+{2\over 3J}\:|\lambda_{xi}|^2  
\nonumber\\ 
&-&\Bigl\{\chi_{xi}\Bigl(\frac38 J\sum_{\sigma} 
f^{\dagger}_{x+i\:\sigma}\:f_{x \sigma} 
+t b^{\dagger}_{x+ i}b_x \Bigr)+\mbox{H.c.}\Bigr\} \nonumber  \\ 
&-& \frac12 
\Bigl\{ \lambda_{xi} 
\Bigl(f^{\dagger}_{x\uparrow}f^{\dagger}_{x+ i\downarrow} 
- f^{\dagger}_{x\downarrow}f^{\dagger}_{x+ i\uparrow}\Bigr)+\mbox{H.c.} 
\Bigr\}\Bigr].   
\label{Hdec1} 
\end{eqnarray} 
In the MFT, the local constraint (\ref{constraint}) is relaxed 
to the following global constraints for averages; 
\begin{eqnarray} 
\langle b^{\dagger}_{x} b_{x} \rangle &=& \delta,\nonumber\\ 
\sum_{\sigma} \langle   f^{\dagger}_{x\sigma} f_{x\sigma} \rangle  
&=& 1-\delta, 
\label{constraint2} 
\end{eqnarray} 
where $\delta\ [\in (0,1)]$ is the doping parameter. 
This modification is
validated when the system is in the CSS state a posteriori because
the CSS implies that the local constraint becomes irrelevant for
holons and spinons. This is the subject of Sect.2B, and some 
supplementary discussion is also given in Sect.6.
Then, $Z(\beta)$ is written as 
\begin{eqnarray} 
Z &=& \int [db][df][d\chi][d\lambda] \:  \exp (-S'),  \nonumber\\ 
S' &=& \int_0^{\beta}d\tau \Bigl[\sum_x \Bigl(  b^{\dagger}_x
\frac{\partial b_x}{\partial\tau}  + 
\sum_\sigma f^{\dagger}_{x\sigma}\frac{\partial f_{x\sigma}}
{\partial\tau} \Bigr) + H_{\rm MF} + H_{\mu} \Bigr], \nonumber\\
H_{\mu} &=& -\sum_{x}\Bigl(\tilde{\mu}_B  b^{\dagger}_{x} b_{x}  
+ \tilde{\mu}_F \sum_{\sigma} f^{\dagger}_{x\sigma} f_{x\sigma} \Bigr), 
\label{ActionS'} 
\end{eqnarray} 
where we replaced $H_{\rm LC}$ by $H_{\mu}$, $\tilde{\mu}_{B,F}$ in 
which are the chemical potentials to enforce (\ref{constraint2}). 
By differentiating the integrand in (\ref{Hdec1}) w.r.t. 
$\chi_{xi}$ and $\lambda_{xi}$, we obtain the following
relations (Schwinger-Dyson equations) among averages;
\begin{eqnarray} 
\langle \chi^{\dagger}_{xi} \rangle &=& \langle
\sum_{\sigma} f^{\dagger}_{x+i\;\sigma}f_{x\sigma} 
+ \frac{8t}{3J}b^{\dagger}_{x+i}b_x \rangle, \nonumber\\
\langle \lambda^{\dagger}_{xi} \rangle &=&  \frac{3J}{2} \langle
 f^{\dagger}_{x \uparrow}f^{\dagger}_{x+i\;\downarrow} 
- f^{\dagger}_{x \downarrow}f^{\dagger}_{x+i\;\uparrow}   \rangle,  
\label{MF}
\end{eqnarray} 
which show that
$\chi_{xi}$ describes hoppings 
of holons and spinons,  while $\lambda_{xi}$ 
describes the resonating-valence-bond (RVB) (NN singlet spin-pair) 
amplitude.

The MFT can be set up first by parametrizing the auxiliary fields as 
\begin{eqnarray} 
\chi_{xi}(\tau) &=& \chi_i U_{xi}(\tau),\ \  
U_{xi}(\tau) \equiv \exp(i A_{xi}(\tau)),\nonumber\\  
\lambda_{xi}(\tau) &=&  \lambda_i V_{xi}(\tau), \ \  
V_{xi}(\tau) \equiv \exp(i B_{xi}(\tau)).  
\label{MFs} 
\end{eqnarray} 
by ignoring the site- and $\tau$-dependence of their amplitudes, 
and then ignoring the  
phase fluctuations by setting $A_{xi} = 0, \ B_{xi} = 0$. 
The effects of these phases are considered in Section 2B. 
Numerical studies were performed for this MFT
in 2D  with various patterns of  the MF's $\chi_i$ and $\lambda_i$. 
One of the most interesting case is the so called 
unform RVB, in which $\chi_i$  are uniform, $\chi_i \equiv \chi$,  
and  $\lambda_i$  are the  d-wave configuration,
$\lambda_1 = -\lambda_2 \equiv \lambda$. 
In Sect.3C, we shall study the general pattern  
of $\lambda_i$  in terms of the GL theory and show the  
stability of this d-wave configuration. 
The MF phase diagram in $\delta$-$T$ plane can be calculated 
in a straightforward manner.   
In Fig.\ref{pd} we show the result of Ref.\cite{IMS}.
There appears several critical temperatures.
Below $T_{\chi}$,  $\chi$ develops. $\chi$ is estimated at small  
$\delta$'s as 
\begin{eqnarray} 
\chi 
&\simeq& \frac{4}{\pi^2}\sin^2 
\Bigl(\frac{\pi}2\sqrt{1-\delta}\Bigr)  
+ \frac{8t}{3J}\delta, 
\label{chiestimate} 
\end{eqnarray} 
at $T_{\rm SG} < T$ where $\lambda_{xi} = 0$.\cite{chiestimate} 
From (\ref{Hdec1}) and (\ref{MF}), 
one expects holons and spinons may hop independently for $\chi \neq 0$.
Thus, in the level of MFT, 
the CSS state is realized below $T_{\chi}$. However, as we shall 
see in Sect.2B, the effect of phase fluctuations of $\chi_{xi}$ 
reduces this CSS onset  temperature down to $T_{\rm CSS}$ which 
is much lower than $T_{\chi}$.
Below  $T_{\rm SG}$, $\lambda$ develops, so the system enters
into the spin-gap state.
There is another transition temperature $T_{\rm BC}$
below which bose condensation of holons, $\langle b_x \rangle \neq 0$,
takes place. To obtain a nonvanishing $T_{\rm BC} > 0$,
a weak three-dimensional coupling  $\chi_3 = \alpha \chi $ is included
in Ref.\cite{IMS}. In the SB MFT, the superconducting phase is 
described by the simultaneous condensations of $\lambda$ and 
$ \langle b_x \rangle $,\cite{supercharge} 
because the order parameter of 
superconductivity, the pairing amplitude of NN hole states, 
is expressed as follows;
\begin{eqnarray} 
&&\langle C_{x \downarrow }C_{x+i \uparrow }
- C_{x \uparrow }C_{x+i \downarrow }\ \rangle \nonumber\\
&=& \langle f^{\dagger}_{x \uparrow } f^{\dagger}_{x+i \downarrow }
- f^{\dagger}_{x \downarrow }f^{\dagger}_{x+i \uparrow } \rangle 
\times \langle b_x  b_{x+i} \rangle 
\nonumber\\
&=& \lambda  \langle b_x \rangle^2.  
\label{superorder} 
\end{eqnarray} 
Thus $T_c(\delta)$ is the lower temperature of $T_{\lambda}(\delta)$ 
and $T_{\rm BC}(\delta)$.

 
\subsection{Beyond MFT: Phase Fluctuations and U(1) Gauge Theory} 
 
In any MFT, it is indispensable to examine the stability of its 
solution. 
Especially, the SB t-J model (\ref{ActionS}) possesses a time-dependent 
U(1) local gauge symmetry,
\begin{eqnarray} 
b_x(\tau) &\rightarrow& e^{i\theta_{x}(\tau)}b_x(\tau), \nonumber\\ 
f_{x\sigma}(\tau) &\rightarrow& e^{i\theta_{x}(\tau)}f_{x\sigma}(\tau)
\nonumber\\
\alpha_x(\tau) &\rightarrow&  \alpha_x(\tau) 
+ \partial_\tau \theta_{x}(\tau).
\label{gaugesym1} 
\end{eqnarray} 
The Lagrange multiplier $\alpha_x(\tau)$ 
is regarded as a time component of the gauge field. 

In the partition function (\ref{ActionS'}) of the decoupled MFT, we
fixed the gauge to the temporal gauge by setting $\alpha_x(\tau) = 0$. 
A careful and precise treatment of this gauge fixing   
is given in Ref.\cite{IMconstraint}, which 
assures us to rederives the results of Papers I. 
After this gauge fixing,
there still remains in the decoupled Hamiltonian (\ref{Hdec1})  
a time-independent residual gauge symmetry 
(with $\tau$-independent $\theta_x$);
\begin{eqnarray} 
b_x(\tau) &\rightarrow& e^{i\theta_{x}}b_x(\tau),  \nonumber\\  
f_{x\sigma}(\tau)  &\rightarrow&  e^{i\theta_{x}}f_{x\sigma}(\tau),
\nonumber\\
\chi_{xi}(\tau) &\rightarrow& e^{-i\theta_{x}}\chi_{xi}(\tau)
e^{i\theta_{x+i}}, 
\nonumber\\
\lambda_{xi}(\tau)  &\rightarrow&  e^{i\theta_{x}}
\lambda_{xi}(\tau)e^{i\theta_{x+i}}. 
\label{gaugesym2} 
\end{eqnarray} 
The last two relations, which are naturally understood
when one recalls (\ref{MF}),
show that the phase  degrees of freedom of the auxiliary fields 
$\chi_{xi}$ and $\lambda_{xi}$ can be regarded as two kinds of gauge 
fields;
\begin{eqnarray} 
A_{xi}(\tau) &\rightarrow& A_{xi}(\tau) -\theta_{x} +\theta_{x+i},
\nonumber\\
B_{xi}(\tau) &\rightarrow& B_{xi}(\tau) + \theta_{x} + \theta_{x+i}, 
\label{A&B}
\end{eqnarray} 
both of which are associated with the common U(1) gauge symmetry.
Thus, the problem of stability around MFT, $A_{xi} = B_{xi} = 0$,
reduces to the dynamics of the resultant gauge theory of
$A_{xi}$ and $B_{xi}$. 
 
Let us first consider the region $T > T_{\rm SG}$, where
$\lambda_{i} = 0$ and one needs to consider only $A_{xi}(\tau)$
or $U_{xi}(\tau)$.
The region  $T < T_{\rm SG}$, where both $A_{xi}(\tau)$ and
$B_{xi}(\tau)$ appear, shall be considered in Sect.2C.
The dynamics of $U_{xi}(\tau)$ is governed by an effective action 
that is obtained by integrating out the holon and spinon variables. 
This idea is similar to that for the $O(N)$ nonlinear $\sigma$-model, 
in which an effective action or a potential of the auxiliary 
(Lagrange multiplier) field is calculated by integrating out 
the original scalar fields $n_{xa}$ satisfying the local constraint
$\sum_{a=1}^{N} n_{xa} n_{xa} =1$. The successive large $N$ analysis
fot 3D system reveals the exitence of a phase in which  the 
local constraint is irrelevant. This phase corresponds to
the CSS phase of the t-J model.\cite{onmodel}
What is important in integrating over $b_{x}$ and
$f_{x\sigma}$ is to intact the dynamics of $U_{xi}$.
For example, if one reduces the lattice system into a simple
continuum field theory at this stage, the action may contain only
quadratic terms of $A_{xi}$ such as the kinetic term, 
$|\partial_{\tau}b(x)|^2$ and $|(\partial_i - i A_i)b(x)|^2$,
and the Maxwell term, $(\partial_i A_j - \partial_j A_i)^2$. 
This action fits well to a perturbative analysis in which  $A_i$ is 
assumed as a small quantity, but it necessarily loses the possibility 
to describe the confinement phase in which $A_i$ is large, although 
the integration over $b(x)$ and $f_{\sigma}(x)$ can be done exactly.
As we proposed in Papers I, we can perfom the integration over $b_{x}$ 
and $f_{x\sigma}$ by an approximate but a nonperturbative method, 
i.e., the hopping expansion in powers of $U_{xi}$, 
which gives rise to an action that is capable to
describe both confinement and deconfinement phases. 
This action $S_{U}$ generally contains two different kinds of terms; 
the electric one $S_{UE}$ that controls the $\tau$-dependence of 
$U_{xi}(\tau)$, and the magnetic one $S_{UM}$ that controls the 
spatial-variation of $U_{xi}(\tau)$. 
Typical forms of them are given as   
\begin{eqnarray} 
S_U &=& S_{UE} + S_{UM}, \nonumber\\
S_{UE}&=&  {1\over g^2_E(T)}\sum_{xi} \int^\beta_0 d\tau  
\partial_{\tau}  U^{\dagger}_{xi}(\tau)\partial_{\tau} 
U_{xi}(\tau),   \nonumber   \\ 
S_{UM}&=& -{1\over g^2_M(T)}\sum_{\rm pl}\prod_{\ell=1}^4 
\int^\beta_0 d\tau_\ell \prod_{\rm pl} U_{xi}(\tau_\ell), 
\label{actions} 
\end{eqnarray} 
where $\prod_{\rm pl} U_{xi}$ denotes the product of four $U_{xi}$'s 
along an oriented plaquette; $U_{xi}U_{x+i,j}U^{\dagger}_{x+j,i}
U^{\dagger}_{xj}$. $g_E(T)$ and $g_M(T)$ are the $T$-dependent 
effective coupling constants. In Papers I, their $T$-dependences 
are calculated explicitly as  
\begin{equation} 
g_E^2(T) \propto T^3, \;\; \; g^2_M(T) \propto const,
\label{coupling} 
\end{equation} 
by using the hopping expansion w.r.t. $U_{xi}$'s.

In the conventional lattice gauge theory,\cite{LGT} 
the action in the temporal gauge $A_{x0} = 0$ is given by 
\begin{eqnarray} 
S^{\rm LGT} &=& S^{\rm LGT}_E+S^{\rm LGT}_M,  \nonumber\\ 
S^{\rm LGT}_E &=& {1 \over e^2} \sum_{xi} \int^\beta_0 d\tau  
\partial_{\tau}  U^{\dagger}_{xi}(\tau)\partial_{\tau} 
U_{xi}(\tau),   \nonumber   \\ 
S^{\rm LGT}_M &=& -{1\over e^2} \sum_{\rm pl} \int^\beta_0 d\tau 
\prod_{\rm pl}  U_{xi}(\tau) \nonumber \\ 
&=& -{1\over e^2} \sum_{\rm pl} \int^\beta_0 d\tau 
\prod_{\rm pl} \cos F_{xij}(\tau), 
\label{LGT} 
\end{eqnarray} 
where $F_{xij}(\tau) (\equiv A_{xi} + A_{x+i\;j} - A_{x+j\;i} 
- A_{xj})$ is the magnetic field penetrating the plaquette. 
In the Hamiltonian formalism,\cite{LGT} 
the corrsponding Hamiltonian $H^{\rm LGT}$ is derived as
\begin{eqnarray} 
H^{\rm LGT} &=& H^{\rm LGT}_E+H^{\rm LGT}_M,  \nonumber\\ 
H^{\rm LGT}_E &=&  e^2 \sum_{x,i} E_{xi}^2, \nonumber\\ 
H^{\rm LGT}_M &=&  -{1\over e^2} \sum_{x,i < j} ( U_{xi} 
U_{x+i,j} U^{\dagger}_{x+j,i} U^{\dagger}_{xj} + \mbox{H.c.}).
\label{HLGT} 
\end{eqnarray} 
Here, the electric field
$E_{xi} (\equiv \partial_{\tau} A_{xi})$ is the variable 
canonically conjugate to $A_{xi}$, so 
the following  uncertainity principle holds;
\begin{eqnarray} 
[ E_{xi}, A_{yj} ] &=& i \delta_{xy}\delta_{ij}, \ \ \ 
\Delta E_{xi}\ \Delta A_{xi} > 1.
\label{uncertainity} 
\end{eqnarray} 
Since $A_{xi} (\in [0, 2\pi])$ is an angle variable, 
$E_{xi}$ takes an integer in its diagonal representation.
When one applies $U_{xi}$ to a state, a segment of electric flux
is created on $(x,x+i)$; $U_{xi}$ is a creation operator of electric
flux. 
The time-independent residual gauge symmetry restricts the physical 
states so as to satisfy the Gauss's law,
\begin{eqnarray} 
\sum_i \nabla_{i} E_{xi} &\equiv & \sum_i \Bigl( 
E_{x+i,i} - E_{xi} \Bigr) = 0,
\label{Gauss} 
\end{eqnarray} 
where $\nabla_i$ is the lattice difference operator;
$\nabla_i f_x \equiv f_{x+i} - f_x$.

The system  (\ref{LGT} - \ref{Gauss}) is known to exhibit 
a CDPT at finite $T$. This was shown by Polyakov and 
Susskind\cite{CDPT}  
by mapping the system  (\ref{HLGT}) to a XY spin model. 
Monte Carlo simulations confirm the CDPT.
Below we present a general explanation why the CDPT takes place. 
For this purpose, let us first ignore the magnetic term 
$H^{\rm LGT}_M$ in (\ref{HLGT}), the reason of which shall 
be explained later.
The partition function is then written as
\begin{eqnarray} 
Z_{\rm LGT} &=& \prod_{xi}\sum_{E_{xi} = -\infty}^{\infty} 
\prod_{x}\delta_{\sum{}_i \nabla_{i} E_{xi}, 0}\ 
\exp(-\beta e^2 \sum_{xi}E_{xi}^2).\nonumber\\
\label{ZLGT} 
\end{eqnarray} 
For large coupling $e^2$ and/or at 
low $T$ such as $e^2\beta \gg 1$,  configurations with small 
$ | E_{xi} |$ are favored.  To investigate the gauge dynamics,  
let us introduce two test charges with $Q = \pm e$ at
$x_1, x_2$. Then the Gauss's law is modified to $\nabla_{i} 
E_{xi} = Q_x \equiv e\delta_{x,x_1} - e\delta_{x,x_2}$, 
and the configuration with the lowest energy is the one with  
electric flux $E_{xi} = \pm 1$ formed along the straight 
line connecting $x_1$ and $x_2$,
and $E_{xi} = 0$ elsewhere. See Fig.\ref{flux}a.
The potential energy of this configuration 
is linear rising, $V(x_1, x_2) = e^2 | x_1 - x_2|$,
so the two charges are confined. In this confinement state, 
$\Delta E_{xi} = 0$, so the relation (\ref{uncertainity}) 
implies $\Delta A_{xi} = \infty$, that is, $A_{xi}$ fluctutes 
violently. On the other hand, for small coupling $e^2$ and/or at 
high $T$ such as $e^2\beta \ll 1$, $E_{xi}$ fluctuates violently, 
$\Delta E_{xi} \gg 1$, which implies $\Delta A_{xi} \ll 1$. Thus,
the fluctuations of $A_{xi}$ is small, and can be treated by usual
perturbation theory. The potential energy  can be evaluated
via exchange of a massless gauge boson, leading to the Coulomb 
potential, $V(x_1, x_2) = e^2 | x_1 - x_2|^{-1}$ in 3D space. 
This is the 
deconfinement state. The electric fluxes are attached
isotropically to each external charge. See Fig.\ref{flux}b.
These considerations  lead us to  a certain finite temperature  
$T_{\rm CDPT}$ at which the CDPT takes place. For fixed $e^2$, 
the string tension 
vanishes and charged particles are deconfined {\it above} 
$T_{\rm CDPT}$. The quark-gluon plasma in QCD at high $T$ 
is such a state. 
Below $T_{\rm CDPT}$, particles with nonvanishing charges that couple
to $A_{xi}$ are confined to form charge-neutral objects. Mesons and 
baryons in QCD are such objects.

Let us estimate the effect of the magnetic term $H^{\rm LGT}_M$.
Let us consider a state  in the confinement phase, $E_{xi}$ of
which takes definite values for all links.
When $H^{\rm LGT}_M$ is applied to this state, each segment of 
electric flux at $(x, x+i)$ is deformed to the one  lying
on a detour $(x, x+j), (x+j, x+j+i), (x+j+i,x+i)$. Thus 
$H^{\rm LGT}_M$ lets $E_{xi}$ to fluctuate and increases 
$\Delta E_{xi}$, and so acts to favor the deconfinement phase. 
Thus $H^{\rm LGT}_M$ may decreases $T_{\rm CDPT}$ certainly, but
$T_{\rm CDPT}$ does not vanish, as the strong coupling expansion 
w.r.t. $ e^{2} \gg 1 $ at $T = 0$ assures us the existence of 
the confinement phase.\cite{LGT}

Let us return to the present case (\ref{actions}) of the t-J model. 
The condition that the deconfinement state occurs may be estimated 
as $\beta g_E^2 < 1$. Due to the 
$T$-dependence of $g_E^2(T)$ as shown in (\ref{coupling}),  
the deconfinement state, hence the CSS, takes place {\it below} 
certain critical temperature $T_{\rm CSS}$;
\begin{eqnarray} 
\beta\ g_E^2  &\simeq& C\  \frac{1}{T}\ T^3 \simeq
\left(\frac{T}{T_{\rm CSS}}\right)^2 < 1.  
\label{Tcss} 
\end{eqnarray} 
For $T <T_{\rm CSS}$,  the MFT works as a first approximation
because $\Delta A_{xi}$  is small there.  
We stress that the reason why our result
$T < T_{\rm CSS}$ for the CSS state is opposite to that of
(\ref{actions}), $T > T_{\rm CDPT}$ as in QCD, is due to the nontrivial 
$T$-dependence of coupling constants (\ref{coupling}), 
which are generated by integrating out holons and spions. 
In Ref.\cite{Nagaosa}, Nagaosa studied certain dissipative gauge 
theory of fermions and obtained the result that a deconfinement 
state appears {\it above} certain temperature or the strength of 
dissipation is strong enough. However, we note that he does not
discuss the possibility that the parameters of his model may be
effective and $T$-dependent.

At first, the above conclusion may seem rather strange since 
the original gauge coupling is infinite (the coefficient of Maxwell 
term is zero). 
Our study shows the possibility that a coupling to matter-fields 
strongly influences the phase structure. In fact,
this fact is now well-established in the elementary particle physics, 
especially in the lattice gauge theory. 
A good example is the non-Abelian gauge model without matter fields,
which is always in the  confinement phase in 3D at $T=0$  
regardless of the strength of the gauge coupling. 
Studies of the lattice gauge theory\cite{iwasaki}, however, 
show that, when  
light $N_f$-flavor quarks are coupled, the phase structure 
drastically changes depending on the value of $N_f$. 
Even at infinite-gauge-coupling limit, this quark-gluon system 
is in a deconfinement phase for $N_f>7$ at $T=0$. 
This example is closely related with the present $U(1)$ gauge-theory 
model for the t-J model. 
First, Polyakov\cite{polyakov} showed that the pure compact $U(1)$ 
lattice gauge system on a 3D lattice without matter fields is always 
in a confinement phase. (This system corresponds to the
2D Hamiltonian system at $T = 0$. In certain approximation,
it can be also regarded as an effective system of a quantum 
system in 3D space at finite $T$. For more details, see 
the paragraph above (\ref{villain}).)
Next, inclusion of nonrelativistic fermions (the spinons in the 
SB representation) can be regarded as inclusion of many-flavor 
light Dirac fermions because of the Fermi surface (line) instead 
of the ``Fermi point" of the Dirac fermion. 
Thus, what we showed for the t-J model in Papers I is nothing but 
the similar deconfining phonomenon induced by ample light fermions. 

\begin{figure} 
  \begin{picture}(200,200) 
    \put(0,0){\epsfxsize 200pt \epsfbox{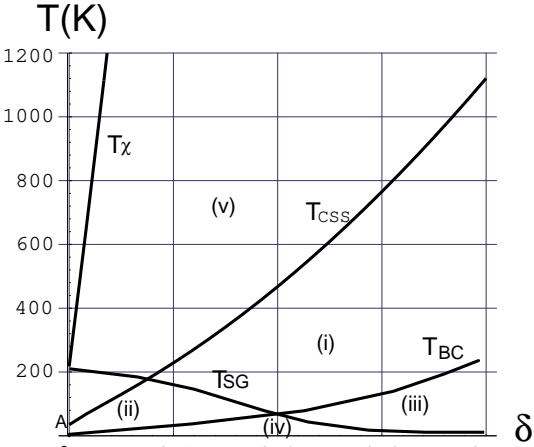}} 
  \end{picture} 
  \caption{ 
    Phase diagram of the SB t-J model in the $\delta - T$ plane. 
Along  $T_{\chi}$ and $T_{\rm RVB}$, $\chi$   and 
$D$ vanish, respectively. $T_{\rm BC}$ is the 
onset $T$ of  Bose condensation, $\langle b_x\rangle \neq 0$. 
$T_{\rm CSS}$ is the critical temperature
{\it below} which the CSS takes place.  
There are five phases: (i) Strange Metal Phase:  
Deconfinement-Coulomb;    
(ii) Spin-Gap Phase:   Deconfinement-Higgs with $D \neq 0$;  
(iii) Fermi-Liquid Phase:  Deconfinement-Higgs  
with $\langle b_x\rangle\neq 0$;    
(iv) Superconducting Phase: Deconfinement-Higgs  
with $D,  \langle b_x\rangle  \neq 0$;  
(v) Electron Phase: Confinement {\it above} $T_{\rm CSS}$.   
} 
\label{pd} 
\end{figure}

\begin{figure} 
  \begin{picture}(240,100) 
    \put(0,0){\epsfxsize 240pt \epsfbox{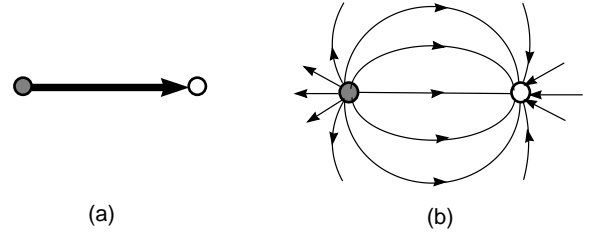}} 
  \end{picture} 
  \caption{ 
Illustration of electric fluxes connecting two external charges.
(a) Confinement state. (b) Deconfinement state. 
    } 
\label{flux} 
\end{figure}

Let us make a couple of comments on the explicit curve 
of $T_{\rm CSS}(\delta)$ in Fig.\ref{pd}.
As mentioned in Sect.2.A,  it is much lower than the MF value 
$T_{\chi}$.
This implies that the effect of gauge fluctuations are significant.
In contrast with Fig.\ref{pd},
the observed $T_{\rm CSS}(\delta)$  seems  decreasing
as $\delta$ is increased.\cite{sato} In Ref.\cite{IMS2},
we observe that the two-body
interactions introduced there tend  to lower $T_{\rm CSS}$ 
for larger $\delta$.
We also note that the value of $T_{\rm CSS}(\delta)$
in Fig.\ref{pd} is not reliable at very small $\delta$ , 
because the long-range AF order in 3D at $\delta\ 
(\leq 0.04)$ 
upsets the validity of the hopping expansion we used.\cite{imcss}
(This point is related with the argument given in Sect.2D below.)

 
\subsection{Spin Gap State} 
 
At low temperatures below $T_{\rm SG}$, the MFT 
of Sect.2 predicts that the NN spin-pairing amplitude $\lambda$ 
develops in addition to $\chi$. This phase is called the spin-gap 
phase, because the spinon excitations in 2D acquire the following 
excitation energy $E(\mbox{\boldmath $k$})$, i.e., an energy gap;  
\begin{eqnarray} 
E^2(\mbox{\boldmath $k$})  
&=&  \left(\frac{3J\chi}{4}   \sum_{i} \cos k_i-\tilde{\mu}_F\right)^2    
+ \left(\lambda \sum_i (-)^i \cos k_i \right)^2. \nonumber\\ 
\label{spinonenergy} 
\end{eqnarray} 
We note that there are several gapless points in $k$ space,
at which $\cos k_1 = \cos k_2 = 2\tilde{\mu}_F/(3J\chi)$.
The mechanism to generate this gap is similar to that of the famous 
energy gap in the BCS model of conventional superconductivity.
Since the spinon part of the MF Hamiltonian (\ref{Hdec1}) has the  
structure,
$H_f \sim  \epsilon f^{\dagger}_\sigma f_\sigma + \lambda 
(f^{\dagger}_\uparrow f^{\dagger}_\downarrow +
f^{\dagger}_\downarrow f^{\dagger}_\uparrow)$,
diagonalization of $H_f$ by a  Bogoliubov transformation gives rise to
$H_f \sim \sum_{k,\sigma}  E(\mbox{\boldmath $k$}) 
\alpha^{\dagger}_{k\sigma}\alpha_{k\sigma},\ 
E(\mbox{\boldmath $k$}) \simeq [\epsilon^2 + \lambda^2]^{1/2}$.

In this phase, there appear two kinds of gauge fields $A_{xi}$ 
and $B_{xi}$ of (\ref{MFs}) and (\ref{A&B}).
Their gauge dynamics can be studied by the action
$S_{UV}$ that is obtained by integrating over  
$b_x$ and $f_{x\sigma}$  by the hopping expansion w.r.t. 
$U_{xi},\ V_{xi}$. 
$S_{UV}$ is calculated in Papers I. It takes the form;
\begin{eqnarray} 
S_{UV} &=& S_{UVE} + S_{UVM}, \nonumber\\
S_{UVE}&=& \int^\beta_0 d\tau  \sum_{xi}  \left[
{1\over g^2_{EU}(T)} \partial_{\tau} U^{\dagger}_{xi}(\tau)
\partial_{\tau} U_{xi}(\tau) \right. \nonumber\\
&+& \left. {1\over g^2_{EV}(T)} \partial_{\tau} V^{\dagger}_{xi}(\tau)
\partial_{\tau} V_{xi}(\tau) \right],   \nonumber   \\ 
S_{UVM}&=& - \int^\beta_0 d\tau  \sum_{x,i<j}   \nonumber\\
&\times & \left[ {1\over g^2_{MU}(T)}
U_{xi}U_{x+i,j} U^{\dagger}_{x+j\,i}U^{\dagger}_{xj}\right. 
\nonumber\\
&+& {1\over g^2_{MV}(T)} 
V_{xi}V^{\dagger}_{x+i\,j}
V_{x+j,i}V^{\dagger}_{xj}\nonumber\\
&+& \left. {1\over g^2_{MUV}(T)} 
V^{\dagger}_{x+i\,j}\,U_{x+j\,i}V_{x\,j}\,U_{x\,i} 
+ {\mbox H.c.}\right], 
\label{actionsUV} 
\end{eqnarray} 
Nature of gauge excitations of this system may be drawed by calculating
the quadratic parts of $S_{UVM}$ in terms of $A_{xi}$, $B_{xi}$.
The  $g^{-2}_{MU} U U U^{\dagger} U^{\dagger}$ term gives rise
to the Maxwell term of  $A_{xi}$. If this term alone is kept, 
it describes a massless  gauge field $A_{xi}$. The term 
$g^{-2}_{MV} V V^{\dagger} V V^{\dagger}$ by itself also describes 
another massless excitations $B_{xi}$.
However, the mixed term $g^{-2}_{MUV}  V^{\dagger} U V U$
generates the term $m_A^2 A_{xi}^2 + m_B^2 B_{xi}^2$, which 
acts as  mass terms of $A_{xi}$ and $B_{xi}$ with 
$m_{A,B}^2 \propto \chi^2 \lambda^2$.
Thus the excitations $A_{xi}$  and $B_{xi}$ become massive 
simultaneously at 
$T < T_{\rm SG}$. This may be called dual Anderson-Higgs mechanism 
for two kinds of gauge fields; one gauge field acts as
a Higgs field for the other gauge field and vise versa.
So we call this phase a Higgs phase, in which the potential 
energy $V(x_1,x_2)$  between two external charges falls off 
exponentially due to exchange of massive
gauge bosons. In Table 1, we compare $V(x_1,x_2)$ for each phase.
The lattice gauge model (\ref{actionsUV}) is complicated, and
no systematic study has been done.
Instead, in Sect.4, we shall study the nature of phase transition at 
$T_{\rm SG}$ in detail using 
the lattice Abelian Higgs model that is derived from the effective 
field theory of Sect.3.

\subsection{Lower Dimensional Cases}

In Sect.2B, we explained that the CSS can be understood as 
a deconfinement phenonenon of the dynamical gauge field $A_{xi}$.
We do not rely explicitly upon the fact that the t-J model of our
interest has a weak but finite three-dimensionality.
Then one may ask whether the same argument can be applied for 
systems in lower spatial dimensions.  We first consider
the 1D case. Then we comment on the 2D case.

For example, the supersymmetric t-J model ($t=J$) in 
one-dimension is known to show the CSS.\cite{1dcss}
However, a straightforward application of the arguments in Papers I,
i.e., the hopping expansion and the Polyakov-Susskind theory of
CDPT \cite{CDPT} to the 1D t-J model leads the result that
only the confinement phase exists.
Technicaly speaking, the Polyakov-Susskind  method
maps the system  to an asymmetric XY spin model \cite{asymmetricxy}
in 1D, and this spin model is 
easily shown to have only a disordered phase, which corresponds to
the confinement phase.\cite{imcss}  
This result seems rather natural because in 1D the Coulomb force 
itself generates a linear-rising confining potential.
There is no possiblity for a deconfinement phase to appear in 1D
if the effective action of the dynamical gauge field is local.
However, if the hopping expansion is not a good approximation or
is not convergent, a deconfinement phase may exist.
This is in sharp contrast to the system in spatial 3D in which 
the existence of a deconfinement phase can be predicted in the 
local gauge action and the hopping expansion.

In order to gain insight into the above problem in 1D and to show 
that a deconfinement phase really exists, we consider gauge dynamics
at low $T$ and  move to
the momentum space instead of the real space assuming the fluctuation
of the dynamical gauge field is small.
This is legitimate since at low $T$ the fluctuation of the gauge 
field becomes small as in the higher-dimensional cases.
Even in this case, the gauge interaction gives rise to linear-rising
confining Coulomb potential if the effects of matters can be ignored.

We shall focus on the spinon contribution for simplicity.
For $U_{xi}=1$, there exist two Fermi points at $\pm k_F$ whose value
depends on the hole concentration.
Then the left and right moving modes can be defined in the usual way.
The continuum field theory is easily obtained and that is nothing but
the 2-flavor massless Schwinger model.
This model can be exactly solved.
Especially by integrating out the spinon or the Dirac fermions,
the effective action of the gauge field is obtained as
\begin{equation}
\int d^2xd^2x'\; A_\mu(x)\Pi_{\mu\nu}(x-x')A_\nu(x')
\label{Pi}
\end{equation}
where $A_0(x)=\alpha_x$ and $U_{x1}=\exp (iA_1(x))$.
The most important term in $\Pi_{\mu\nu}(k)$ is the ``mass term"
of the gauge field,
\begin{equation}
\Pi_{\mu\nu}(k) \propto \Big(\delta_{\mu\nu}
-{k_\mu k_\nu \over k^2}\Big).
\label{Gmass}
\end{equation}
The mass term comes from a ``massless mode" in the 
particle-anti-particle channel, which 
is a direct consequence of the kinematics in 1D and 
behaves like a Higgs boson.
It is obvious that $\Pi_{\mu\nu}(k)$ is finite as $k\rightarrow 0$
and so the higher-order terms of the hopping expansion 
in the real space are to give the main contribution to it.

The above consideration in momentum space assuming small and 
smooth fluctuations 
of the gauge field gives a consistent solution of the deconfinement
phase which is realized as a Higgs phase.
The hopping expansion in the real space cannot uncover this 
possiblity.

Next, let us consider  the 2D case. 
The method of Paper I gives rise to an asymmetric
XY spin model \cite{asymmetricxy} in 2D. 
The MC simulation \cite{janke}
and the analytical method \cite{imcss} show that this model 
has an order-disorder phase transition, which corresponds to 
a CDPT. These analyses show that the CDPT occurs in higher
$T$ than the spin-gap generation temperature $T_{\rm SG}$
as in 3D (See Fig.\ref{pd}). Thus the effective model
is nothing but the 2D (symmetric) XY model, and the CDPT
is  the Kosterlitz-Thouless transition of infinite-order. 
Although the validity of hopping expansion
has not yet been clarified, very existence of a CDPT may survive
in the correct treatment.
There is a special point in 2D. The potential energy between two
slave particles with the opposite charges {\it in Coulomb phase}  
is logarithmic $V(r) \sim \ln (r)$ due to the 2D Coulomb force.
Thus the slave particles are ``confined" even in the Coulomb
phase in the  sense that $\lim_{r \rightarrow \infty} V(r) =
\infty$. This requires a careful study of the nature of
quasiparticles in the Coulomb phase of the pure 2D system.
As mentioned, inclusion of a finite three dimensionality
releases us from this problem. 

 
\section{Effective Field Theory of the Slave-Boson t-J Model} 
\setcounter{equation}{0} 
 
 
\subsection{Hamiltonian of the Continuum Field Theory and 
Integration over Holons and Spinons} 
  
Let us consider the low temperature region $T < T_{\rm CSS}$, 
where the fluctuations of gauge field $A_{xi}$ is small, as
explained in Sect.2B. Thus it is reasonable to derive
the low-energy effective theory there in the form of local gauge
field theory, and apply it the usual perturbation theory to 
evaluate the effect of $A_{xi}$.   
Let us start with the Hamiltonian $H_{\rm MF}$ of 
(\ref{Hdec1}) and translate the lattice variables  
to the continuum fields 
such as  $f_{x\sigma} \to a f_\sigma(\mbox{\boldmath $x$})$ 
by using the lattice constant $a$.
For $U_{xi}$ we set\cite{LGT}
\begin{eqnarray}
U_{xi} &=& \exp(i a A_i(\mbox{\boldmath $x$})) \nonumber\\
&\simeq& 1 
+  ia A_i(\mbox{\boldmath $x$}) -\frac{1}{2} a^2
 A_i(\mbox{\boldmath $x$})^2. 
\end{eqnarray}
The Hamiltonian $H_{c}$ of the continuum field theory is then
calculated as 
\begin{eqnarray} 
  H_c &=&  \int d^2 x\: \Bigl[  \frac{3}{4 a^2} J\chi^2 
  + m_F\:\chi \Bigl( |\lambda_{s}(\mbox{\boldmath $x$})|^2 
    +|\lambda_{d}(\mbox{\boldmath $x$})|^2\Bigr) \Bigr]\nonumber\\ 
  &+&\int d^2 x\: \Bigl[\frac{1}{2m_B} 
    \sum_i\Bigl|D_i b(\mbox{\boldmath $x$})\Bigr|^2 
     -\mu_B\Bigl|b(\mbox{\boldmath $x$})\Bigr|^2\Bigr]\nonumber\\ 
  &+&\int d^2 x\: \Bigl[\frac1{2m_F}\sum_i 
    \Bigl|D_i f_{\sigma}(\mbox{\boldmath $x$})\Bigr|^2 
     -\mu_F\Bigl|f_{\sigma}(\mbox{\boldmath $x$})\Bigr|^2\Bigr] 
\nonumber\\
&+&\int \frac{ d^2 k}{(2 \pi)^2} \Bigl[ 
  \Delta_{\rm SG}(\mbox{\boldmath $k$}) 
  f^{\dagger}_{\uparrow}(\mbox{\boldmath $k$}) 
  f^{\dagger}_{\downarrow}(-\mbox{\boldmath $k$})
  \! +\!\mbox{H.c.} \Bigr], 
\label{Hc}
\end{eqnarray} 
where $\mu_{B, F}$ denote the chemical potentials  
for the continuum theory,  
$f_{\sigma}(\mbox{\boldmath $k$})$ is the Fourier transform of 
$f_{\sigma}(\mbox{\boldmath $x$})$, 
\begin{eqnarray}
f_{\sigma}(\mbox{\boldmath $x$}) &=&
\int \frac{ d^2 k}{(2 \pi)^2} f_{\sigma}(\mbox{\boldmath $k$})
\exp(i\mbox{\boldmath $k$}\mbox{\boldmath $x$}),
\end{eqnarray}
and 
\begin{eqnarray} 
\lambda_{s}(\mbox{\boldmath $x$}) &=& \frac12[\lambda_{1}
(\mbox{\boldmath $x$})+\lambda_{2}(\mbox{\boldmath $x$})],  
\nonumber\\
\lambda_{d}(\mbox{\boldmath $x$}) &=&  \frac12[\lambda_{1}
(\mbox{\boldmath $x$})-\lambda_{2}(\mbox{\boldmath $x$})], 
\nonumber\\
\frac{1}{2m_B} &=& t\chi a^2, \quad
\frac{1}{2m_F} = \frac38 J\chi a^2,\nonumber\\
D_i &=& \partial_i - i A_i,\quad
k_F^2  =  \frac{2\pi}{a^2}(1-\delta) = 2m_F \mu_F.
\end{eqnarray} 
$D_i$ is the covariant derivative w.r.t. 
the gauge field $A_i(\mbox{\boldmath $x$})$, and 
$k_F$ is the Fermi momentum of spinons. 
The last term of $H_c$ in (\ref{Hc}) describes the 
coupling of the spin-gap amplitude to spinon pairs.
Originally, it has the form of
\begin{eqnarray}
  & &\int \frac{ d^2 k\: d^2 q}{(2 \pi)^4} \Bigl[ 
  \Delta_{\rm SG}(\mbox{\boldmath $k$},\mbox{\boldmath $q$}) 
  f^{\dagger}_{\uparrow}(\mbox{\boldmath $k$} 
  \! +\! \frac{\mbox{\boldmath $q$}}2) 
  f^{\dagger}_{\downarrow}(-\mbox{\boldmath $k$} 
  \! +\! \frac{\mbox{\boldmath$q$}}2) 
  \! +\!\mbox{H.c.} \Bigr], \nonumber \\
&& \Delta_{\rm SG}(\mbox{\boldmath $k$},\mbox{\boldmath $q$}) 
\simeq 2(1-\delta)\Bigl(\frac{k^{2}_{1}-k^{2}_{2}}{k^2_F}\Bigr) 
  \lambda_{d}(\mbox{\boldmath $q$}) 
  -2\delta\:\lambda_{s}(\mbox{\boldmath $q$}). 
\label{Deltasg}
\end{eqnarray}
$\Delta_{\rm SG}(\mbox{\boldmath $k$},
\mbox{\boldmath $q$})$ consists of the first 
$d$-wave term and the second $s$-wave term. 
At  {\it half-filling}, $\delta = 0$, the $s$-wave term vanishes.
In $H_c$ of (\ref{Hc}), we simplify $\Delta_{\rm SG}$
by keeping only its  $d$-wave term and ignoring its 
$\mbox{\boldmath $q$}$-dependence as
\begin{equation} 
\Delta_{\rm SG}(\mbox{\boldmath $k$}) 
=  2(1-\delta)\Bigl(\frac{k^{2}_{1}-k^{2}_{2}}{k^2_F}\Bigr) 
  \lambda_{d},  
\label{DelSG} 
\end{equation} 
where $ \lambda_{d}  = \lambda_{d}(\mbox{\boldmath $0$})$ 
($\langle \lambda_i \rangle \simeq (-)^{i} \lambda_{d}$).
We also set $\mbox{\boldmath $q$} = 0$ 
in $f^{\dagger}_{\sigma}$ of (\ref{Deltasg}).

To obtain the effective action $L_{\rm eff}(A_i, \lambda_i)$ of 
$\lambda_i(\mbox{\boldmath $x$}, \tau)$ and 
$A_i(\mbox{\boldmath $x$}, \tau)$,
we integrate over $b$ and $f_{\sigma}$ by the standard bilinear 
integrations in the partition function,
\begin{eqnarray} 
Z &=& \int [db][df][d\lambda_i][dA_i]\exp(-\int d\tau L_c) \nonumber\\
&=& \int [d\lambda_i][dA_i]
\exp(-\int d\tau L_{\rm eff}(A_i, \lambda_i)), \nonumber\\
L_c &=&   \int d^2x ( b^{\dagger}
\partial_{\tau}b  + \sum_{\sigma} f^{\dagger}_{\sigma}\partial_{\tau}
f_{\sigma}  ) + H_{c}.
\end{eqnarray}

 
\subsection{Gauge-Field Propagator} 
 
The propagator of the gauge field, 
$D_{ij}(\mbox{\boldmath $x$}, \tau)$, 
is generated by fluctuations (vacuum polarization) 
of spinons and holons, i.e., 
\begin{eqnarray} 
D_{ij}(\mbox{\boldmath $x$}, \tau) 
&\equiv& \langle A_i(\mbox{\boldmath $x$}, \tau)A_j(0, 0)\rangle  
\nonumber\\ 
  &=& \Bigl[ \Pi_{F}+\Pi_{B}\Bigr]^{-1}_{ij},  \nonumber\\
\Pi_{F,B\;ij}(\mbox{\boldmath $x$}, \tau)
&\equiv& - \langle J_{F,B\;i}(\mbox{\boldmath $x$},\tau)  
J_{F, B\;j}(0,0)\rangle_{\rm 1PI} 
\nonumber\\ 
&+&\delta_{ij}\delta(\mbox{\boldmath $x$})\delta(\tau)n_{F,B},  
\end{eqnarray} 
where 
\begin{equation}
n_F = \frac{1-\delta}{a^2}, \ \ \ n_B = \frac{\delta}{a^2},
\end{equation}
are the holon and spinon density and
$\Pi_{F,B\;ij}$ represent one-particle-irreducible (1PI) 
diagrams of spinon and holon loops made of the following
currents  coupled to $A_{i}$;
\begin{eqnarray} 
J_{{F}i} &\equiv&  
\frac{1}{2m_{F}}\sum_{\sigma}(i\bar{f}_{\sigma} 
\partial_i f_{\sigma}+{\rm H.c.}), \nonumber\\  
 J_{Bi} &\equiv& \frac{1}{2m_{B}}(i\bar{b}\partial_i b
+{\rm H.c.}).  
\end{eqnarray} 
In the Coulomb gauge, $\partial_i A_i = 0$, the propagator 
at momentum  $\mbox{\boldmath $q$}$ and  Matsubara frequency 
$\epsilon_l \equiv 2\pi l /\beta$ is written as    
\begin{eqnarray} 
D_{ij}(\mbox{\boldmath $q$}, \epsilon_l)&\equiv&
\int \frac{d^2 q}{(2\pi)^2} \int \frac{d\tau}{\beta}
e^{-i(\mbox{\boldmath $qx$}+\omega_l \tau)}
D_{ij}(\mbox{\boldmath $x$}, \tau) \nonumber\\
&=&  
\left(\delta_{ij}-\frac{q_i q_j}{q^2}\right) 
D(\mbox{\boldmath $q$}, \epsilon_l), 
\nonumber\\ 
  D(\mbox{\boldmath $q$}, \epsilon_l) &=&  
\Bigl[\Pi_{F}(\mbox{\boldmath $q$}, \epsilon_l) 
+\Pi_{B}(\mbox{\boldmath $q$}, \epsilon_l)\Bigr]^{-1}. 
\label{d} 
\end{eqnarray} 
 
Since we shall need $D$ later in  calculating $\rho$,  
we calculate $D$ below 
in the random-phase approximation as  
\begin{eqnarray}
D &\simeq & \Bigl[\Pi^R_B + \Pi^R_F\Bigr]^{-1}.  
\end{eqnarray}
When the spin gap is sufficiently small, its effect to $\Pi^R_F$ 
 is evaluated by perturbation theory. We obtain  
\begin{eqnarray} 
  &&\frac{\Pi^{R}_F(q, \epsilon_l)}{g^2}\simeq\left\{ 
\begin{array}{rl} 
      \frac{q^2}{12\pi m_F} 
        +\sqrt{\frac{2 n_F}{\pi}}\frac{|\epsilon_l|}{q} 
        +\frac{n_F^S(T)}{m_F}, 
        & |\epsilon_l|\ll v_Fq\\ 
      \frac{n_F}{m_F}, 
      & |\epsilon_l| \gg v_Fq 
\end{array} 
\right. , \nonumber\\
\label{pirf}   
\end{eqnarray} 
where we used the relation, $\mu_F\simeq\pi n_F/m_F$ 
and $v_F\equiv k_F/m_F$. 
The second term $\propto |\epsilon_l|/q$ in the first line
represents the dissipation due to fermions. This term appears
also in the studies of non-Fermi liquids and $\nu = 1/2$
quantum Hall effect.\cite{dissipation}
$n_F^S(T)$ 
is the superfluid density of spinons and  
is calculated  for small 
$\Delta_{\rm SG}(\mbox{\boldmath $k$})/(k_B T)$ as  
\begin{eqnarray} 
  n_F^S &\simeq& \frac{n_F}{2\pi}\int d\phi  
\left|\frac{\Delta_{\rm SG} (\mbox{\boldmath $k$})}{2k_BT}\right|^2 
\nonumber\\
&=& \frac{n_F}{2}(1-\delta)^2
\left(\frac{\pi \lambda}{2k_BT}\right)^2 
\label{sfd} 
\end{eqnarray} 
with $k_1 /k_2 = \tan \phi$  
and $|\mbox{\boldmath $k$}| = k_F$. 
We note that the  last constant term of (\ref{pirf}) represents  
a mass term of $A_i(\mbox{\boldmath $x$},\tau)$.  
$\Pi^R_B$ is calculated as  
\begin{eqnarray} 
  \frac{\Pi^{R}_B(q, \epsilon_l)}{g^2 f_B(|\mu_B|)}&\simeq&\left\{ 
    \begin{array}{rl} 
      \frac{q^2}{24\pi m_B} 
        +\sqrt{\frac{\tilde{n}_B(T)}{\pi}}\frac{|\epsilon_l|}{q}, 
        & |\epsilon_l|\ll \tilde{v}_B(T)q \nonumber\\  
      \frac{\tilde{n}_B(T)}{m_B}, 
      & |\epsilon_l| \gg \tilde{v}_B(T)q 
    \end{array} 
  \right. , \nonumber\\
\label{pirb} 
f_B(\epsilon)&\equiv&\frac{1}{\exp(\beta \epsilon)-1},
\nonumber\\ 
\tilde{n}_B(T)&\equiv&  \frac{n_B}{f_B(|\mu_B|)},\ \ 
\tilde{v}_B(T)\ \equiv\  \frac{\sqrt{4\pi \tilde{n}_B(T)}}{m_B}. 
\end{eqnarray} 
Eqs.(\ref{pirf}) and (\ref{pirb}) are obtained for small 
$q (\ll \pi/a)$.  
For large $q$'s, they should be replaced by 
anisotropic expressions due to   
$\Delta_{\rm SG}(\mbox{\boldmath $k$})$. 
These anisotropy can be ignored as long as the spin gap is 
sufficiently small. 
 
 
\subsection{Ginzburg-Landau Theory with Gauge Field} 
 
The most dominant contributions to $Z$ from the integrations over 
$\lambda_i$ and $A_i$ come from their static ($\tau$-independent) 
modes. It is seen 
for $A_i$ by (\ref{pirf},\ref{pirb}). 
So we keep only the static ($\tau$-independent)
modes in the effective Lagrangian density $L_{\rm eff}$.
The partition function is then written as
\begin{eqnarray}
Z &=& \int \prod_{\mbox{\boldmath $x$}}\prod_i
d\lambda_i(\mbox{\boldmath $x$})
dA_i(\mbox{\boldmath $x$})\ \exp(-\beta\int d^2x\ L_{\rm eff}).
\label{zzeromode}
\end{eqnarray}
$L_{\rm eff}$ is given up to the fourth-order 
in fields and derivatives by 
\begin{eqnarray} 
L_{\rm eff}&=&  c\:\sum_i\Bigl( 
2\delta^2|{\cal D}_{i}\lambda_{s} |^2 
+(1-\delta)^2|{\cal D}_{i}\lambda_{d} |^2 
\Bigr)   \nonumber\\ 
&+& c\: \delta(1-\delta) \Bigl( 
\overline{{\cal D}_{1}\lambda_{s}} 
{\cal D}_{1}\lambda_{d} 
-\overline{{\cal D}_{2}\lambda_{s}} 
{\cal D}_{2}\lambda_{d} + \mbox{H.c.} 
\Bigr)  \nonumber\\ 
&+& \frac1{12\pi\bar{m}}\sum_{i j}\frac14 F_{ij} F_{ij}
+V(\lambda_s,\lambda_d), \nonumber\\
V(\lambda_s,
\lambda_d)&=& 
a_s |\lambda_{s}|^2 + a_d |\lambda_{d}|^2\nonumber\\ 
&+& 
4 b\: \delta^4|\lambda_{s}|^4 
+\frac{3}{2} b\ (1-\delta)^4|\lambda_{d}|^4 
\nonumber\\ 
&+& 2 b\:\delta^2(1-\delta)^2 
\Bigl(4|\lambda_{s}|^2 
|\lambda_{d}|^2 + \bar{\lambda}_{s}^2 
\lambda_{d}^2 
+ \bar{\lambda}_d^2 \lambda_{s}^2 \Bigr),\nonumber\\
\label{GLpotential}
\end{eqnarray} 
where the bars represents complex conjugate quantities.
The GL coefficients, etc. are given by 
\begin{eqnarray} 
\frac{1}{\bar{m}}  &\equiv& \frac{1}{m_F} + 
\frac{f_B(|\mu_B|)}{2m_B}, 
\nonumber\\
a_s &=& m_F\chi-\frac{2}{\pi}m_F\delta^2 
\ln\Bigl(\frac{2e^{\gamma}}{\pi}\beta\omega_\lambda\Bigr),
\nonumber\\ 
a_d &=& 
m_F\chi - \frac{m_F}{\pi}(1- \delta)^2\ln\Bigl(\frac{2e^{\gamma}}
{\pi} \beta\omega_\lambda\Bigr), 
\nonumber\\ 
b &=& \frac{7\zeta(3)}{8\pi^{3}} \beta^2 m_F, 
\quad c = \frac{7\zeta(3)}{32\pi^{3}m_F} \beta^2  k_F^2,
\nonumber\\ 
{\cal D}_{i} &=& \partial_{i}-2iA_{i}, \quad 
F_{ij} = \partial_{i}A_{j} - \partial_{j}A_{i}, 
\end{eqnarray} 
and $\gamma$ is the Euler number.    
$\omega_\lambda$ is the cutoff of the spinon  energy  
[$\xi \equiv k^2/(2m_F) - \mu_F$] in the one-loop integrals  
representing spinon pairings, 
and is estimated as $\omega_\lambda \sim O(\mu_F)$.
 
The system favors the d-wave state 
at small $\delta$'s, and the s-wave state at  large $\delta$'s. 
Actually, the transition temperature $T_{\lambda_{d(s)}}$
at which $\lambda_{d(s)}$ starts to condense is estimated 
from the potential energy $V(\lambda_s,\lambda_d)$ of 
(\ref{GLpotential}) by setting
$a_{\lambda_{d(s)}} = 0$, and one sees that $T_{\lambda_d}
>\ T_{\lambda_s}\ (T_{\lambda_d}
<\ T_{\lambda_s})$ for small (large) $\delta$'s.
Let us consider the region of small $\delta$'s and 
focus on the $d$-wave component by setting $\lambda_s = 0$
and parameterizing as 
\begin{eqnarray}
\lambda_1(\mbox{\boldmath $x$}) &=& 
\lambda \exp[ i\phi(\mbox{\boldmath $x$})], \ \  
\lambda_2(\mbox{\boldmath $x$}) = 
-\lambda \exp[i\phi(\mbox{\boldmath $x$})]. 
\label{phi}
\end{eqnarray}
Here we introduced the MF $\lambda$, 
the $d$-wave spin-gap amplitude, for the radial parts of  
$\lambda_{i}(\mbox{\boldmath $x$})$, ignoring their fluctuations. 
Then  we have the  effective Lagrangian density, 
\begin{eqnarray} 
L_{\rm eff} & = & L_{\lambda} + L_A, \nonumber\\ 
L_{\lambda}& =& 
a_d  \lambda^2 + \frac{3}{2} b\: (1-\delta)^4\lambda^4,\nonumber\\ 
L_A & =&  c\ (1-\delta)^2\lambda^2 
\Bigl(\partial_{i}\phi - 2A_{i}\Bigr)^2  
 +    \frac1{12\pi\bar{m}} 
\sum_{i j}\frac14 F_{ij} F_{ij}. \nonumber\\
\label{Leff} 
\end{eqnarray} 
The MF result is obtained by minimizing the above $L_{\lambda}$ 
w.r.t. $\lambda$ as 
\begin{eqnarray} 
k_{\rm B} T_{\rm SG} 
&=&\frac{2e^{\gamma}}{\pi}\: \omega_\lambda 
\exp\Bigl[-\frac{\pi\chi}{(1-\delta)^2}\Bigr] 
\nonumber\\ 
c\:(1-\delta)^2\lambda^2
&\simeq& \frac{k_F^2}{12\pi m_F} 
\Bigl(1-\frac{T}{T_{\rm SG}}\Bigr), 
\label{MFT} 
\end{eqnarray} 
where $T_{\rm SG}(\delta)$ is the critical temperature below  
which $\lambda$ develops. (See $T_{\rm SG}$ of Fig.1.)
The second result holds for $T$ near $T_{\rm SG}$. 
We assume a small but finite three-dimensionality to stabilize these 
MFT results. This point and other 
details of the transition at $T_{\rm SG}$ are studied in Sect.4.


\subsection{Gauge-Field Mass in  Variational Method } 
 
Let us consider the effect of $L_A$ upon $L_{\lambda}$  
by integrating over $A_i$. 
We have treated $A_{i}$ as a noncompact gauge field 
in spite of being originally compact. 
This procedure is appropriate in respect of the kinetic term of $A_{i}$, 
because we consider the region $T < T_{\rm CSS}$. 
In short, the compactness of the gauge-field kinetic term
should be respected in studying the CDPT at $T_{\rm CSS}$, 
while the compactness
of the coupling to the Higgs field is crucial to study
the transition into Higgs phase at $T_{\rm SG}$. 
So we  respect the compactness of the mass term by
replacing $(A_i - \partial_i\phi/2)^2$ in $L_A$
to the periodic one. 
Then the new Lagrangian $L_B$ which replaces $L_A$ is expressed as 
\begin{eqnarray} 
L_B &=&\frac{1}{12\pi\bar{m}} 
\sum_{i j}\frac14 F_{ij} F_{ij} 
\nonumber\\ 
&+&  c (1-\delta)^2\lambda^2 \cdot 
\frac1{a^2} 
\Bigl[4-\sum_{i}2\cos\Bigl(2 a B_{i}\Bigr)\Bigr], 
\end{eqnarray} 
where we introduced  
the Proca (massive vector) field $B_i$,
\begin{eqnarray}
B_i  &\equiv& A_i - \frac{1}{2}\partial_i\phi, \nonumber\\ 
F_{ij} &=& \partial_i B_j - \partial_j B_i.
\end{eqnarray} 
Noe that if the cosine term in $L_B$ is expanded up to the 
second order, $L_A$ is recovered. Let us take the unitary gauge
$\phi = 0$. 
Then the path integrals reduce as 
\begin{eqnarray}
\prod_x d\phi\ \prod_{x,i} dA_i &\equiv& 
\prod_x d\phi \prod_{x,i} dB_i 
\rightarrow \prod_{x,i} dB_i.
\end{eqnarray}

Let us estimate the gauge-field mass $m_{A}$ by 
the variational method.\cite{variational} 
We choose the variational Lagrangian $L_B'$ for $L_B$  as 
\begin{eqnarray} 
L_B' &=&\frac{1}{12\pi\bar{m}} \Bigl( 
\sum_{i j}\frac14 F_{ij} F_{ij} 
+\sum_i\frac{m^2_{A}}2\: B_i B_i 
\Bigr), 
\end{eqnarray} 
where $m_{A}$ is introduced here as a variational parameter. 
There holds the Jensen-Peierl inequality for the free energy 
density $F^{\rm exact}_B$ of $L_B$,
\begin{eqnarray}
&&F^{\rm exact}_B       \le F^{\rm var}_B \equiv
   F_B' +\frac{1}{V}\int d^2x
   \langle L_B - L_B' \rangle',\nonumber\\
&&\exp(-\beta F^{\rm exact}_B V) = \int \prod_{x,i} dB_i 
\exp\Bigl(-\beta \int  d^2x L_B\Bigr), 
\nonumber\\
&&\exp(-\beta F_B' V)  = \int \prod_{x,i} dB_i 
\exp\Bigl(- \beta \int d^2x L_B'\Bigr), \nonumber\\
&&\langle O \rangle' \equiv \int \prod_{x,i} dB_i O 
\exp\Bigl(- \beta \int d^2x L_B' + \beta F_B' V\Bigr),\nonumber\\
\label{variational}
\end{eqnarray}
where $V = \int d^2 x $ is the system volume.
By  taking the definition of the propagator at the same point and 
the trace of a functional operator $\hat{O}$ as 
\begin{eqnarray} 
  \langle B_i(\mbox{\boldmath $x$}) B_j(\mbox{\boldmath $x$})\rangle 
&\equiv&\lim_{\mbox{\boldmath $y$}\to \mbox{\boldmath $x$}} 
\langle B_i(\mbox{\boldmath $x$}) B_j(\mbox{\boldmath $y$})\rangle, 
\nonumber\\ 
\mbox{Tr}\: \hat{O} 
&\equiv&\int d^2 x \lim_{\mbox{\boldmath $y$}\to \mbox{\boldmath $x$}} 
\langle \mbox{\boldmath $x$}| \hat{O} | \mbox{\boldmath $y$}\rangle, 
\end{eqnarray} 
the variational free energy density $F^{\rm var}_B$ is calculated by  
integrating over $B_i$ as follows; 
\begin{eqnarray} 
F^{\rm var}_B(m_A) 
&=& F_B(0) +\frac{k_{\rm B} T}{8\pi}\:m^2_{A} \nonumber\\
&-&\frac{4 c\: (1-\delta)^2\lambda^2}{a^2} 
\Bigl(\frac{m^2_A}{q^2_c}\Bigr)^{\frac{T}{\Theta}}, 
\nonumber\\ 
k_{\rm B}\Theta&\equiv& 
\frac1{3 a^2 \bar{m}}=\chi\Bigl[\frac{J}4+\frac{t}3\:
f_B(-\mu_B)\Bigr], 
\label{freeenergy} 
\end{eqnarray} 
where $q_c$ is the momentum cutoff of the $B_i$ modes. 
Its explicit value is estimated later in (\ref{qc}). 
$\Theta $ is a function $\Theta(\delta,T) $ of $\delta$ and $T$.
In $F^{\rm var}_B$ of (\ref{freeenergy}) we have omitted 
higher-order terms of 
$O ( m^4_A/q^4_c)$ since we are interested in small $m_A^2/q^2_c$.
This assumption is justified a posteriori by the result itself.
 
By minimizing $F^{\rm var}_B(m_A)$ w.r.t. $m_A$, we obtain 
\begin{eqnarray} 
m^2_A & = & 
q^2_c\Bigl[\frac{96\pi \bar{m}}{q_c^2} 
\:c\:(1-\delta)^2\lambda^2\Bigr]^{2 d}, \nonumber\\ 
d &=& \frac{1}{2}\frac{\Theta}{\Theta -T}.
\end{eqnarray} 
We note that the fluctuations of $A_i$ do not affect 
the MF result (\ref{MFT}) as long as $d > 1$ 
since the corrections then becomes higher-order than $\lambda^4$. 
Thus the  gauge-field  mass $m_A$ at fixed $\delta$
starts to develop continuously at $T_{\rm SG}(\delta)$ as 
\begin{eqnarray} 
m_A(\delta,T) &\propto& (T_{\rm SG}(\delta) 
- T)^{d(\delta,T_{\rm SG}(\delta))}. 
\end{eqnarray} 
That is the exponent $d$ is neither $1/2$ nor a constant, 
and drastically changes especially when $T_{\rm SG} \sim T_A$, 
where $T_A$ is a root of the equation $T_A=\Theta(\delta,T_A)$, 
at which $d(\delta,T)$ diverges. 
This is in strong contrast with the noncompact case, in which 
the effect of $A_i$ upsets the MF result of GL Lagrangian 
$L_{\lambda}$  as stated. Actually, in the calculation of 
Ref. \cite{Halperin&Lubensky&Ma}, 
$L_A$ in 3D generates a notorious $\lambda^3$ term.

 
\section{Phase Transition into the Spin Gap State: 
Lattice Abelian Higgs Model} 
\setcounter{equation}{0} 
 
In this section, we clarify the nature of the
phase transition at $T_{\rm SG}$ described by the effective field 
theory (\ref{Leff}) of Sect.3C.
The reader who is interested in the practical applications
of the effective field theory may skip this section, and
directly go to Sect.5.

To study the transition at $T_{\rm SG}$, it is useful to introduce
a lattice model that is suggested by the effective field theory
(\ref{Leff}), because ample methods of analysis are available for 
lattice models.
The most natural lattice model for this purpose is a noncompact 
version of the so called lattice Abelian Higgs model, which
is defined by
\begin{eqnarray} 
Z_{\rm AH} &=& \prod_{x} \int_{-\pi}^{\pi} d\phi_{x} 
\prod_{x,i}\int_{-\infty}^{\infty} dA_{xi} 
\exp(-S_{\rm AH}), \nonumber\\
S_{\rm AH}&=&-\rho \sum_{x,i}  
\cos [\nabla_i\phi_x-  A_{xi}] 
+{\kappa \over 2}\sum_{x, i < j}\Big(\epsilon_{ij}\nabla_i 
A_{xj}\Big)^2,\nonumber\\
 \rho &\propto& \lambda^2, \; \;
\kappa \propto 1/\bar{m},
\label{ZAH} 
\end{eqnarray} 
where $\phi_x\in [-\pi,\pi]$ is the phase of a Higgs field on $x$  
corresponding to $\phi(\mbox{\boldmath $x$})$ of (\ref{phi})
and $A_{xj} \in [-\infty, \infty]$ is a noncompact gauge field   
corresponding to  $A_i(\mbox{\boldmath $x$})$.
The two parameters $\rho$ and $\kappa$ are determined by
the parameters of the effective  model of Sect.3. 
(This $\rho$ should not be confused with the dc resistivity 
although we use the same latter.)
$S_{\rm AH}$ has the  local U(1) gauge symmetry under
\begin{eqnarray} 
\phi_x &\to& \phi_x + \theta_x, \nonumber\\
A_{xi} &\to& A_{xi} + \nabla_i \theta_x,  
\label{ZAHGI} 
\end{eqnarray} 
Its naive continuum limit reduces to (\ref{Leff}). 

A similar but better  studied model is the 
compact version of (\ref{ZAH}) \cite{savit,ES} defined by
\begin{eqnarray} 
Z^{\rm c}_{\rm AH} &=& \prod_{x} \int_{-\pi}^{\pi} d\phi_{x} 
\prod_{x,i}\int_{-\pi}^{\pi} dA_{xi} \exp(-S^c_{\rm AH}), \nonumber\\
S^{\rm c}_{\rm AH}&=&-\rho \sum_{x,i}  
\cos [\nabla_i\phi_x-qA_{xi}] 
-\kappa \sum_{x, i < j}\cos(\epsilon_{ij}\nabla_i A_{xj}).
\nonumber\\
\label{ZAHC} 
\end{eqnarray} 
It is obtained from (\ref{ZAH})
by (i) letting $A_{xj} \in [-\pi, \pi]$, (ii) replacing
the quadratic Maxwell term $(\epsilon_{ij}\nabla_i A_{xj})^2$
by the periodic (compact) one, 
$\cos(\epsilon_{ij}\nabla_i A_{xj})$, and (iii) introducing $q$. 
$q$ is the charge of the Higgs field, and the phase structure
may differ for each $q$. On the contrary, 
in the noncompact case of (\ref{ZAH}), we have set $ q= 1 $ since
$q$ can be eliminated  by rescaling $A_i \to q^{-1} A_i,\ \kappa 
\to q^2 \kappa$ and so irrelevant there. 

The main difference between these two models on 3D lattice
is that  monopoles (which are nothing but instantons in the 
3D lattice system \cite{polyakov})
are allowed in (\ref{ZAHC}), while  not in (\ref{ZAH}). 
A monopole is a particular configuration of $A_i$,
\begin{equation} 
\nabla_i H_i  =  4\pi \delta_{x,o}, \; \; 
H_i \equiv \epsilon_{ijk} \nabla_j A_k,
\label{instanton} 
\end{equation}
which is a solution of the field equation and has the 
nonvanishing divergence of ``magnetic field" $H_i$. 
Condensation of these monopoles let $A_i$ random,
$\Delta A_i = \infty$, and drive the system
into the confinement phase.\cite{polyakov} 
Since  we  are interested in $T$ around 
$T_{\rm SG}$  $( < T_{\rm CSS})$ which is in the deconfinement
phase, we consider the noncompact model (\ref{ZAH}).
On the other hand, we respect the periodicity (compactness) 
of the first term in $S_{AH}$ of (\ref{ZAH}), which plays the 
important role in the transition at $T_{\rm SG}$.
This point was already pointed out in Sect.3D.
At the end of this section, we shall mention the compact model
(\ref{ZAHC}).

Let us first consider the isotropic 3D model (\ref{ZAH}) with
$i = 1,2,3$, because the phase structure of the  
quasi-2D t-J model in which we are interested is governed by the  
three dimensionality of its effective lattice model. (The effect 
of anisotropy is discussed later in the paragraph containing 
eq.(\ref{quasi2D}) below.) Here we comment on the dimensionality
of an effective lattice model. It is well known that the path integral
quantization maps a quantum system in $D$-dimensional space to
a classical system in $D+1$ dimensional space, where the extra one 
dimension represents the imaginary time, $\tau \in (0,\beta)$.
In many cases, this $D+1$ classical system is approximated
by an effective model in $D$-dimensional space. A good example
is nothing but our model, $L_{\rm eff}$ of (\ref{zzeromode}),
where only the $\tau$-independent modes appear.
As long as $T \neq 0$, the extra dimension has  a finite
extension $\beta$, and so it should not affect the long-range
properties and the phase structure.
(The case $T=0$ is different of course.)
 The $T$ dependence
are incorporated into the coefficients of the effective action.
The reduction to the $D$-dimensional effective model is reasonable
in this sense. Because the Abelian Higgs model is originated
from $L_{\rm eff}$ of (\ref{zzeromode}), both models should be 
considered in the same spatial dimensionality.

Let us modify the gauge-Higgs term of $S_{\rm AH}$ 
to the following Villain (periodic Gaussian) form; 
\begin{eqnarray} 
&& \exp\Big[\beta\cos(\nabla_i\phi_x-A_{xi})\Big] \nonumber\\
 & \to & 
e^{\beta}\sum_{n_{xi}= -\infty}^{\infty} 
\exp\Big[-{\beta \over 2}(\nabla_i\phi_x-A_{xi}+2\pi n_{xi})^2\Big]
\nonumber\\
&= &e^{\beta}\sum_{m_{xi}= -\infty}^{\infty} 
\exp\Big[-{2 \over \beta}m_{xi}^2  + i m_{xi}
(\nabla_i\phi_x-A_{xi}) \Big].
\label{villain}
\end{eqnarray} 
This replacement keeps the periodicity (compactness) in $\phi_x$. 
It is then straightforward to perform 
a duality transformation\cite{savit}   
to reach the following dual representation;
\begin{eqnarray} 
&&Z_{\rm AH}  =  \prod_{x i}{\sum_{J_{xi}}}'
\int_{-\infty}^{\infty} dC_{xi} \times \nonumber\\
&& \exp\Big[-\sum_{x i} {1 \over 4\rho} 
\Big((\epsilon_{ij\ell}\nabla_jC_\ell)^2 + m^2 C_i^2 \Big)  
+2\pi i \sum_{xi} J_{xi} C_{xi}\Big] \nonumber \\ 
&&=  Z_0   \prod_{x i} {\sum_{J_{x i}}}'  
\exp\Big(- 4\pi^2\rho \sum_{x i}\sum_{y j}
J_{xi}D_{ij}(x-y;m^2) J_{yj}\Big),  \nonumber\\
\label{Zdual} 
\end{eqnarray} 
where $m^2=\rho/\kappa$.  The dual variables
$J_{xi}\ ( = 0, \pm1, \pm2,...)$ and 
$C_{xi}\ (\in [-\infty, \infty])$ are defined on the link
$(x,x+i)$ of the dual lattice (which is obtained from the original
lattice by shifting each site $x$ to $x + (1+2+3)/2$).
From  the saddle point equations, one obtains the relation,
\begin{eqnarray} 
\langle J_{xi} \rangle &\propto& 
\langle \epsilon_{ijk} \nabla_j (\nabla_k \phi_x - A_{xk}) 
\rangle\nonumber \\
\label{vortex} 
\end{eqnarray} 
Thus $J_{xi}$ represents the gauge-invariant "vortex current",
because $ \vec{J} \propto \vec{\nabla} \times \vec{\nabla}\phi$
for $\vec{A} =0$. (Recall that the vorticity
in the continuum is defined by $\oint d\phi$ along a closed line. 
On the  lattice, a vortex with a  phase change $2\pi$
along a plaquette in the 12 plane 
is expressed by $J_{x3} = \pm 1$, and so on.)
From the gauge invariance, $J_{xi}$ can be chosen so as to
conserve at each site,\cite{savit,ES}
\begin{eqnarray} 
\sum_i \nabla_i J_{xi} &=& 0.
\label{conserve} 
\end{eqnarray} 
The prime on $\sum_{J_{xi}}$ of (\ref{Zdual}) implies 
this condition. The vector field $C_{xi}$ mediates the force 
between two vortex current elements, $J_{xi}$, and generates
the potential $D_{ij}(x-y;m^2)$ as shown in the last line of 
(\ref{Zdual}).

$Z_0$ is the partition function of the vector field 
$C_{xi}$ with mass $m$, 
the Green's function of which is given as  
\begin{eqnarray} 
&& Z_0 = \exp(-\frac{1}{2}{\rm Tr Log} D_{ij}(x-y;m^2)),\nonumber\\
&& \langle C_{xi} C_{yj} \rangle = D_{ij}(x-y;m^2)\nonumber\\
&&= \Big[\delta_{ij}-{\nabla_i\nabla_j  
\over m^2}\Big] D(x-y;m^2),  \nonumber   \\ 
&& (-\nabla^2+m^2)D(x-y; m^2)=\delta_{xy}. 
\label{latticeD} 
\end{eqnarray} 

The system (\ref{Zdual}) is characterized by the density of
vortices. If there are no votices, (\ref{vortex}) shows
that every $\phi_x$ is stabilized to a fixed value, say
$\phi_x = 0$, with $\Delta \phi_x = 0$, and
gives rise to a coherent condensation of  the Higgs field
with a long-range order, $\langle \lambda_x \rangle  \neq 0$. 
Appearance of vortices let $\phi_x$ fluctuate, and the 
prolifiration of vortices, i.e., a vortex condensation,  leads to 
destruction of the long-range order of the Higgs field.
Thus one expects a phase transition between these two states;
(i) the Higgs phase where the Higgs field condenses 
and vortices do not condense,
and (ii) the Coulomb phase where vortices condense and 
no Higgs condensation, $\langle \lambda_x \rangle = 0$.
A valance  between the energy and the entropy determines 
the critical line. 
From (\ref{Zdual}), the energy of a vortex current
with the strength $| J_{xi} | = 1$ and the length $L$ 
is estimated as  
\begin{equation} 
E(L) \sim 4\pi^2 \rho D(0;m^2) L, 
\label{vortexE} 
\end{equation} 
whereas its entropy $S(L)$ is estimated as 
\begin{equation} 
S(L) \sim \ln \mu^L, 
\label{vortexEnt} 
\end{equation} 
with some numerical constant $\mu \sim 4.7$ for the simple cubic 
lattice.\cite{ES}  Thus we estimate the transition temperature
$T^*$ (which should be identified with $T_{\rm SG}$) as 
\begin{eqnarray} 
F &=& E(L) - T S(L) \simeq 0, \nonumber\\
T^*  &\simeq&  4\pi^2 \rho D(0;m^2)\ \ln \mu.
\label{Tsg} 
\end{eqnarray} 
$\rho D(0;m^2)$ in $T^*$ is estimated as
\begin{eqnarray} 
\rho \; D(0;m^2)  &=&
\rho \int^\pi_{-\pi}d^3p  
{1\over \sum_i \sin^2p_i+m^2}\nonumber\\
&\simeq& \left\{ 
\begin{array}{cr} 
    \rho, \  & \rho/\kappa \ll 1  \\
\kappa,   \     &  \rho/\kappa \gg 1
\end{array} 
\right. .
\label{rhoD} 
\end{eqnarray}   
The duality between the Higgs field $\lambda(x)$ 
and the vortex density $v(x)$ is illustrated in Table 2.
 
Detailed studies of the above system have been 
performed by numerical simulations.\cite{Higgs} 
There the radial degrees of freedom of the Higgs field
are treated also as dynamical variables. 
It is found that the phase transiton really occurs, 
and the  order of the transition depends on the parameters 
of the model.
For a small Higgs-four-point interaction, it is of first-order,
whereas for large four-point interaction, it becomes a 
``continuous" transition.\cite{Kleinert} 
The mass of the gauge boson was also calculated in detail 
near the continous  phase transition as a function of the 
Higgs mass. In the isotropic $3$D case, the gauge boson mass 
develops as $(T_{\rm SG} - T)^{1/2}$ as 
predicted by the MFT of (\ref{MFT}), i.e., 
$m_A \propto \lambda$ and $d=1/2$. 
 
For the quasi-2D case, a similar phase transiton between 
the Higgs phase and the Coulomb phase is expected to occur, 
i.e., the phase structure of the quasi-2D system is governed  
by its three dimensionality. 
This expectation is quite natural from a general view point of 
universality in renormalization group.
The effect of the three dimensionality may be
described by introducing anisotropy parameters 
in the Hamiltonian such as 
\begin{eqnarray}
H&=&-\sum_{x, i,\sigma}t_i \Bigl(b^{\dagger}_{x+ i}
f^{\dagger}_{x \sigma} 
f_{x+ i\:\sigma}b_x+\mbox{H.c.}\Bigr)\nonumber\\ 
&-& \sum_{x, i} \frac{J_i}{2} \Bigl| 
f^{\dagger}_{x\uparrow}f^{\dagger}_{x+ i\downarrow} 
- f^{\dagger}_{x\downarrow}f^{\dagger}_{x+ i\uparrow}\Bigr|^2,
\nonumber\\
t_1 & =& t_2 \equiv t,\ \ t_3 = \alpha t, \nonumber\\
J_1 & =& J_2 \equiv J,\ \ J_3 = \gamma \alpha J.
\label{quasi2D}
\end{eqnarray}
The parameter $\alpha$ controls the quasi-two dimensionality, 
i.e., $\alpha=1$ corresponds to the isotropic 3D case and  
$\alpha=0$ to the pure 2D case.
Another parameter $\gamma$ respects the difference
in $t_i$ and $J_i$. If we use the relation $J = 4t^2/U$ obtained
in the derivation of the t-J model from the Hubbard model, 
we have $\gamma = \alpha$.

As an explicit example of the related quasi-2D models, 
we have studied
the quasi-2D antiferromagnetic Heisenberg model at finite 
$T$.\cite{YTIM}
By using the quasi-2D O(3) nonlinear $\sigma$ model, we
showed that the critical temperature $T_{\rm AF}$ behaves as 
$T_{\rm AF} \sim 1/|{\rm log}\alpha|$ for small $\alpha$,
which vanishes for 2D as expected, and the correlation functions
exhibit 3D critical behabior in the vicinity of the transition 
point $T_{\rm AF}$ with the interval of 
$\Delta T \sim O(1/({\rm log}\alpha)^2)$.

Yet another example is the the quasi-2D XY spin model.\cite{JM}
There the critical temperature in the 2D case remains finite, 
at which a phase transition of Koterlitz-Thouless type takes place.
We note that the application of the 
Polyakov-Susskind theory\cite{CDPT} to the effective lattice gauge 
theory of Sect.2B in 2D case maps the system into the 2D XY 
spin model, hence predicts a   
phase transition of Kosterlitz-Thouless type 
at {\it finite} $T_{\rm SG} $ for 2D. 
This point is different from the Heisenberg model, for which
the critical temperature vanishes as $\alpha \rightarrow 0$. 

Returning to the present quasi-2D model (\ref{ZAH}), 
a phase transition should occur at $T = T_{\rm SG} > 0$,
which may be either continuous or discontinuous.  
In the very narrow interval $\Delta T$ around $T_{\rm SG}$, 
the system exhibits a genuine 3D critical properties. 
This interval, however, may be too small to access by the experiments. 
Outside of this $\Delta T$, the system should exhibit the 2D behaviors,
which can be studied well by using the results obtained in Sect.2D.
Analytical studies and the MC simulations of 
the quasi-2D case of (\ref{ZAH}) are welcome
to confirm these expectations. 
 
Let us now comment on the compact U(1) Abelian Higgs model 
(\ref{ZAHC}) in 3D.
The duality transformation can be performed as in 
the noncompact case.\cite{ES}
The resultant partition function is given by the same expression as  
(\ref{Zdual}) but  the vortex current $J_{xi}$ can terminate 
at locations of instantons (monopoles) and anti-instantons. 
Thus eq.(\ref{conserve}) is modified to
\begin{equation} 
\nabla_i J_{xi} = q Q_x,
\label{instantonsource} 
\end{equation} 
where $Q_x \ (= 0, \pm1, \pm2, ...)$ is the instanton density.
The partition function is obtained by summing over all
the instanton configurations $\sum_{Q_x}$ in addition to 
$\sum_{J_{xi}}$.
Existence of sufficient instantons converts the Coulomb phase 
into the confinement phase. 

Recently, this 3D compact $U(1)$ gauge-Higgs model 
was examined in the continuum field theory by Nagaosa and
Lee \cite{NLdual} to study the competition of Bose condensation
versus gauge-field fluctuations. They employed  a duality 
transformation similar to the one above on a lattice. 
They concluded that, in 3D 
the vortex condensation always takes place for $q=1$, and no
phase transitions occur; the system is
always in the confinement phase. 
Their argument does not depend on details of the model
and the same conclusion is reached for any systems in which
vortices and instantons exist and a single vortex couples to 
an instanton.
However, their conclusion contradicts the numerical 
studies \cite{Higgs} 
which exhibit a genuine phase transition from the confining 
phase to the Higgs phase. 
In Appendix, we shall revisit this problem and explain
why the argument given in Ref.\cite{NLdual} is incorrect. 
There is a critical line in the $\kappa-\rho$ plane, and
this line terminates at some point.

Finally, we mention the recent work of Balent et al.\cite{BFN}.
They studied strongly-correlated electron systems in spatial 2D  
and argued its phase strucutre by mapping the electrons into 
bosons by using the Chern-Simons gauge theory
and then by applying the duality transformation (See Table 2).
They also introduced two composite fields for spin and charge 
degrees of freedom as in the bosonization in 1D case. 
One may expect a similar method may be used for the spinons 
in the SB t-J model. However, their way to derive the low-energy
theory of the composite bosons is accompanied with certain
assumptions and not straightforward. 
Anyway, we should stress here that their approach to the  
strongly-correlated 2D system is very close to that to the 1D 
system for which the bosonization separates the spin and charge 
degrees of freedom quite easily, but loose the ability to
describe the Higgs phase (i.e., $T_{\rm SG} = 0$). 
On the other hand, in our approach, we take the three 
dimensionality of the high-$T_c$ cuprates seriously, which
is crucial to predict the realistic phase structure with 
$T_{\rm SG} > 0$ as we explained above. 

In the following section, we shall calculate the conductivity 
in the spin-gap phase. There we assume the 
quasi-two-dimensionality and the genuine phase transition
at $T_{\rm SG}$. 
However,  we checked that the kinematical formulae 
appearing in our calculations, which we obtained
assuming  $T_{\rm SG} > 0$, have  smooth and weak dependence
on $\alpha$ near $\alpha = 0$. From this observation and the 
estimation $\alpha \sim 10^{-5}$,\cite{YTIM} we set 
$\alpha = 0$ in the calculations in Sect.5 for simplicity.
The errors caused by this 
simplification do not modify the qualitative conclusion there. 
 

\section{Resistivity} 
\setcounter{equation}{0}

In Sect.5A, we rederive the Ioffe-Larkin formula for
conductivity with a general assignement of EM charges 
of holons and spinons. In Sect.5B, the detailed expression
of the conductivity is obtained. In Sect.5C, the spin gap effect
on the conductivity is calculated by using the gauge-boson mass
$m_A$ of Sect.3D.
 
\subsection{Linear Response Theory and Ioffe-Larkin Formula} 
\begin{figure} 
  \begin{picture}(240,100) 
    \put(0,0){\epsfxsize 240pt \epsfbox{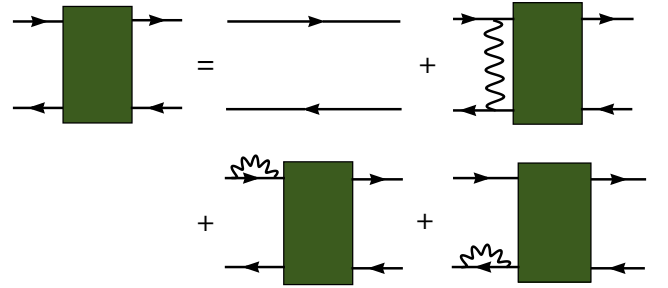}} 
  \end{picture} 
  \caption{ 
Graphical representation of the Schwinger-Dyson equation for 
$\tilde{\Pi}_{F(B)}$. 
    } 
\label{SD} 
\end{figure}

Here we shall consider the response of the system  
to the external EM field $A^{\rm ex}_{i}\ 
(i = 1,2,3)$. We shall work in the temporal gauge, 
where $A_0 = 0,\ A^{\rm ex}_0 = 0$. 
In principle, the effective action of $A^{\rm ex}_{i}$ 
is obtained by integrating out all the 
quantum fields (spinon, holon, dynamical gauge filed $A_{i}$) 
and one can calculate the response to $A^{\rm ex}_{i}$ 
from that action. 
There is an ambiguity in the way how $A^{\rm ex}_{i}$  
couples to spinons and holons.  Namely, one may assign
the EM charge of a holon, $Q_h$  and the charge
of a spinon, $Q_s$ arbitrarily as long as the charge of an 
electon $e ( < 0)$ is expressed as 
\begin{eqnarray}
 e = Q_s - Q_h,
\label{charge}
\end{eqnarray}
as suggested by the relation (\ref{slaveboson}). 
This problem was investigated first by Ioffe and Larkin\cite{il} 
and the result is that this ambiguity does not affect the
expectation values of gauge-invariant physical quantities. 
For the conductivity, this can be seen explicitly 
in the formula given below.
Let us recall that the relevant part of the Hamiltonian
with general charges $Q_h,\ Q_s$ is written as
\begin{eqnarray} 
  H_{\rm em} &=&\frac{1}{2m_B} \int d^2 x
    \sum_i\Bigl|(\partial_i -iQ_h A^{\rm ex}_i -iA_i)
     b(\mbox{\boldmath $x$})\Bigr|^2 \nonumber\\ 
  &+ &\frac1{2m_F} \int d^2 x \sum_i 
    \Bigl|(\partial_i -iQ_s A^{\rm ex}_i -iA_i) 
    f_{\sigma}(\mbox{\boldmath $x$})\Bigr|^2
\nonumber\\ 
  &+ &\int \frac{ d^2 k}{(2 \pi)^2} \Bigl[ 
  \Delta_{\rm SG}(\mbox{\boldmath $k$}) 
  f^{\dagger}_{\uparrow}(\mbox{\boldmath $k$}) 
  f^{\dagger}_{\downarrow}(-\mbox{\boldmath $k$}) 
  \! +\!\mbox{H.c.} \Bigr].
\label{Hem}
\end{eqnarray} 
Note that the last line
acquires no explicit $A^{\rm ex}_{xi}$ dependence since
its  expression  in  coordinate space $\sim \int dx^2
\Delta_{x i} f^{\dagger}_{x \sigma} f^{\dagger}_{x+i,-\sigma}$
remains unchanged under the EM gauge transformation,
$\Delta_{x i} \rightarrow \Delta_{x i}
\exp(i(\theta^{\rm ex}_x +\theta^{\rm ex}_{x+i})),\ 
f_{x \sigma} \rightarrow  f_{x \sigma}
\exp(i\theta^{\rm ex}_x)$. 
By integrating out the spinon and holon fields,  
we obtain the following effective Lagrangian; 
\begin{eqnarray} 
L_{\rm eff}[A_{i}, A^{\rm ex}_{i}] 
&=&  
(A_{i}+  Q_s A^{\rm ex}_{i}) \Pi_F^{ij} 
(A_{j}+  Q_s A^{\rm ex}_{j}) \nonumber\\
&+& (A_{i} + Q_h A^{\rm ex}_{i}) \Pi_B^{ij}
(A_{j} +  Q_h A^{\rm ex}_{j}) 
+ O(3), \nonumber\\
\end{eqnarray} 
where $O(3)$ represents terms higher than quadratic in $A_{i}$ 
and/or $A^{\rm ex}_{i}$.  
In order to integrate $A_{i}$ concretely, the above 
Lagrangian needs to be approximated by a quadratic one. 
However, the $O(3)$ terms are very important, because 
they renormalize the spinon propagator, the holon propagator, 
and the vertices. 
This effect is partially included by replacing $\Pi_{F(B)}^{ij}$ 
with their renormalized quantities, $\tilde{\Pi}_{F(B)}^{ij}$,
represented by Fig.\ref{SD}. 
With this replacement, the Lagrangian is expressed as
\begin{eqnarray} 
L_{\rm eff}[A_{i}, A^{\rm ex}_{i}] &=& 
 (A_{i} + Q_s A^{\rm ex}_{i}) \tilde{\Pi}_F^{ij}
(A_{j} + Q_s A^{\rm ex}_{j}) \nonumber\\ 
&+& (A_{i} + Q_h A^{\rm ex}_{i}) \tilde{\Pi}_B^{ij}
(A_{j} + Q_h A^{\rm ex}_{j}).
\nonumber 
\end{eqnarray} 
After integrating over $A_i$, we obtain
\begin{eqnarray} 
&&L_{\rm eff}[A^{\rm ex}_i] = 
 A^{\rm ex}_i \tilde{\Pi}_{ij} A^{\rm ex}_j \nonumber\\ 
\tilde{\Pi} &=&
Q_h^2 \tilde{\Pi}_B +Q_s^2 \tilde{\Pi}_F \nonumber\\
&-&(Q_h \tilde{\Pi}_B + Q_s \tilde{\Pi}_F)
(\tilde{\Pi}_B+\tilde{\Pi}_F)^{-1}
(Q_h \tilde{\Pi}_B + Q_s \tilde{\Pi}_F)\nonumber\\
&=&(Q_h -Q_s)^2\Bigl[\tilde{\Pi}_B^{-1} + \tilde{\Pi}_F^{-1}
\Bigr]^{-1}\nonumber\\
&=& e^2 \Bigl[\tilde{\Pi}_B^{-1} + \tilde{\Pi}_F^{-1}
\Bigr]^{-1},
\label{PI}
\end{eqnarray} 
where $\tilde{\Pi}$ is nothing but the response function 
of electrons.  

Finally, from the linear-response theory and $L_{\rm eff}
[A^{\rm ex}_i],$ 
the dc conductivity $\sigma (\equiv \sigma_{11} = \sigma_{22})$  
is expressed as  
\begin{eqnarray} 
&&\sigma_{ij}   
=  \lim_{\epsilon \rightarrow 0}\lim_{q\rightarrow 0} 
\frac{1}{-i \epsilon}\tilde{\Pi}_{ij}
(\mbox{\boldmath $q$},-i\epsilon),  
\nonumber\\
&&\tilde{\Pi}_{ij}(\mbox{\boldmath $q$}, \epsilon)  
= e^2 \Bigl[\tilde{\Pi}_F^{-1}(\mbox{\boldmath $q$}, \epsilon)  
+ \tilde{\Pi}_B^{-1}(\mbox{\boldmath $q$}, \epsilon)
\Bigr]^{-1}_{ij}.  
\end{eqnarray} 
So one arrives at the Ioffe-Larkin formula,
\begin{eqnarray} 
&&\sigma^{-1}  = \sigma_B^{-1} 
+ \sigma_F^{-1},
\end{eqnarray}  
where $\sigma_{B(F)}$ is the conductivity of holons (spinons).
 
The last expression of (\ref{PI}) is symmetric w.r.t. 
the holon contribution $\tilde{\Pi}_B^{-1}$ and the spinon 
contribution $\tilde{\Pi}_F^{-1}$, and depends only on 
the difference (\ref{charge}). The reason why the ambiguity
of charge assignment
\begin{eqnarray}  
Q_h \to Q_h + c\; e, \ \ Q_s \to Q_s + c\; e, 
\label{shift1}
\end{eqnarray}  
disappears in the final expression of the conductivity
is that one can make a shift of the integration variable
\begin{eqnarray}  
A_{i} \rightarrow A_{i} - c\; e A^{\rm ex}_{i}
\label{shift2}
\end{eqnarray}  
so as to eliminate the $c$-dependence from (\ref{Hem}).
By a shift of integration variable, the partition function
and physical quantities are of course unchanged.


\subsection{Expression of Resistivity} 
In the spin-gap state, the spinon conductivity  
diverges $\sigma_F \rightarrow \infty$ due to a superflow 
generated by the spin-gap condensation $\langle
\lambda_{xi} \rangle \neq 0$. 
This is an analog of the well-known fact in the BCS theory 
that the electron resistivity vanishes below $T_c$ due to a 
superflow generated by a Cooper-pair condensation. Actually,  
$A_{\rm eff}^F$ has the same structure as the BCS model.  
Thus the total resistivity   $\rho$
in the spin-gap state is equal to  
the resistivity of holons, 
\begin{eqnarray}
\rho &=& \sigma^{-1} = \sigma_B^{-1} + \sigma_F^{-1}\nonumber\\
&\rightarrow& \sigma_B^{-1}. 
\label{resistivity}
\end{eqnarray}
Effects of spinons to $\rho$ is indirect, but
certainly exist and show up through the dressed propagator 
$D(\mbox{\boldmath $q$}, \epsilon_l)$  
in calculating $\tilde{\Pi}_B$.  
 
Now we calculate the response function $\tilde{\Pi}_B$.  
By solving the Schwinger-Dyson equation approximately 
according to Ref.\cite{oim}, we obtain  
\begin{eqnarray} 
\tilde{\Pi}_{B\;ij}(0,\epsilon_l) &\simeq& -\frac1\beta\sum_n 
\int \frac{d^2q}{(2\pi)^2}\frac{q_iq_j}{{m_B}^2}
R_B(q, \epsilon_n;\epsilon_l) 
\nonumber\\ 
&&\times \ \ \frac{i\epsilon_l}{i\epsilon_l 
-i\epsilon_l\Gamma_B(q, \epsilon_n; \epsilon_l) 
-\Delta\Sigma_B(q, \epsilon_n; \epsilon_l)}, \nonumber\\ 
R_B(q, \epsilon_n;\epsilon_l) 
&\equiv& G_B(q, \epsilon_n)G_B(q, \epsilon_n+\epsilon_l), 
\nonumber\\ 
G_B(q, \epsilon_n) 
&\equiv& \left(i\epsilon_n - \frac{q^2}{2m_B} + \mu_B\right)^{-1}. 
\label{pib} 
\end{eqnarray} 
 
$\Delta\Sigma_B(q, \epsilon_n; \epsilon_l)$, representing  
diagrams containing self-energy of holons $\Sigma_B(q, \epsilon_n)$,  
is necessary to keep gauge invariance, 
\begin{eqnarray} 
\Delta\Sigma_B(q, \epsilon_n; \epsilon_l) 
&\equiv&  R_B^{-1}(q, \epsilon_n;\epsilon_l) 
\times \Bigl[\Sigma_B(q, \epsilon_n)G_B^2(q, \epsilon_n) \nonumber\\ 
&-&\Sigma_B(q, \epsilon_n+\epsilon_l)G_B^2(q, \epsilon_n+\epsilon_l)
\Bigr], 
\end{eqnarray} 
However, in the perturbative calculation, this combination vanishes  
in the dc limit by the symmetry under summations.  
We expect this term dose not contribute to the dc resistivity, 
and neglect it hereafter. 
  
\begin{figure}
  \begin{picture}(210,210)
    \put(0,0){\epsfxsize 200pt \epsfbox{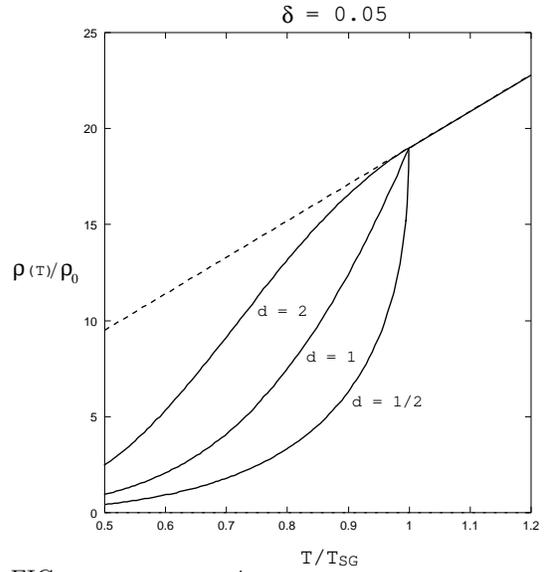}}
  \end{picture}
  \caption{Plot of the resistivity $\rho$ divided by
  $\rho_0\equiv 2\pi m_Fk_B T_{\rm SG}/(e^2n_F)$  for 
  $\hat{\lambda}_{\delta}(T)\simeq \lambda_0(1-T/T_{\rm SG})^d$ 
  $(d = 1/2, 1, 2)$. 
  For definiteness we chose $\delta=0.05$, $\lambda_0 = 
  2k_B T^{*}/ \pi$, $2\pi n_B/m_B = 4\delta k_BT_{\rm SG}/
  (1-\delta)$ and $J/t = 0.35$. 
  }
\label{rhod}  
\end{figure}

$\Gamma_B(q,\epsilon_n;\epsilon_l)$ represents vertex diagrams,  
and contributes to $\sigma_B$. It is given by 
\begin{eqnarray} 
\Gamma_B(q,\epsilon_n;\epsilon_l) 
&\equiv&\left(\frac{g}{m_B}\right)^2\frac1\beta\sum_{n'} 
\int \frac{d^2q'}{(2\pi)^2} \nonumber\\ 
&\times&  
\left\{\frac{\mbox{\boldmath $q$} 
\times (\mbox{\boldmath $q$}'-\mbox{\boldmath $q$})} 
{|\mbox{\boldmath $q$}'-\mbox{\boldmath $q$}|}\right\}^2 
\frac{\mbox{\boldmath $q$}\cdot(\mbox{\boldmath $q$}' 
-\mbox{\boldmath $q$})}{q^2} 
\nonumber\\ 
&\times& D(|\mbox{\boldmath $q$}'-\mbox{\boldmath $q$}|, 
\epsilon_{n'}-\epsilon_n) R_B(q',\epsilon_{n'};\epsilon_l),
\nonumber\\ 
\label{gamma} 
\end{eqnarray} 
where $\mbox{\boldmath $q$}\times\mbox{\boldmath $q$}' 
\equiv q_xq'_y-q_yq'_x$. 
We fix the length of $q$  in  $\Gamma_B$ to a typical 
length $\tilde{q}_B$,  
\begin{eqnarray}
\tilde{q}_B^2 \equiv 4\pi \tilde{n}_B(T).  
\end{eqnarray}
This $\tilde{q}_B$ 
is determined so that the  similar integral, (\ref{pib}) 
with an relaxation time in the holon propagator  
and $i\epsilon_l\Gamma_B+\Delta\Sigma_B \to 0$, 
being evaluated by this approximation, gives rise to the 
correct result. Furthermore, $\tilde{q}_B$ behaves as 
$\tilde{q}_B \sim \sqrt{2 m_B k_BT}$ 
at the temperature region, $k_BT\gg n_B/m_B$. 
So this is a natural choice for the typical momentum scale  
of holons. 
We also fix the length 
of $q'$ of $D$ in eq.(\ref{gamma}) to $\tilde{q}_B$ to obtain
\begin{eqnarray} 
\Gamma_B(\tilde{q}_B,\epsilon_n;\epsilon_l) 
&\simeq&-\frac{g^2{\tilde{q}_B}^2}{8\pi^2m_B}\frac1\beta\sum_{n'} 
    \int^{\pi}_{-\pi}d\phi\;\sin^2\phi 
\nonumber\\   
& \times& D\left(\tilde{q}_B\sqrt{2(1-\cos\phi)}, 
 \epsilon_{n'}-\epsilon_n\right) \nonumber\\ 
& \times& \int^{\infty}_{|\mu_B|}dE\;\frac1{i\epsilon_{n'} -E} 
\frac1{i\epsilon_{n'} +i\epsilon_l -E}.\nonumber\\ 
\label{GammaB}
\end{eqnarray} 
We consider the underdoped region,  
$n_B \ll n_F \ (\delta \ll 1),$  
and temperatures around $ \beta^{-1}\sim n_B/m_B$.  
Assuming that $D(q,\epsilon_l)$ in $\Gamma_B$ dominates in the region 
near the static limit $\epsilon_l = 0$, we use the upper expressions 
in eqs.(\ref{pirf}) and (\ref{pirb}). In the denominator of $D$,  
the dissipation term is larger than the $q^2$ term, 
\begin{eqnarray}
&&\sqrt{\frac{2\bar{n}}{\pi}} \frac{|\epsilon_l|}q >
\frac{q^2}{12\pi \bar{m}}, 
\nonumber\\
&&\sqrt{\bar{n}} \equiv
 \sqrt{n_F} + f_B(|\mu_B|)\sqrt{\frac{\tilde{n}_B(T)}2},
\end{eqnarray} 
as long as  $\epsilon_l \neq 0$, since their ratio is 
$(\tilde{q}_B^2/\bar{m})/(\sqrt{\bar{n} }/(\tilde{q}_B\beta))  
\sim {\cal O}(\sqrt{\tilde{n}_B(T)/n_F})$ and small.  
So the $n'$-sum is dominated at $\epsilon_{n'}=\epsilon_n$. 
Then we get 
\begin{eqnarray} 
\Gamma_B(\tilde{q}_B,\epsilon_n;\epsilon_l) 
&\simeq&-\frac{3 \bar{m}}{4\pi m_B}\frac1\beta 
    \int^{\pi}_{-\pi}d\phi\frac{\sin^2\phi} 
{(1-\cos\phi)+\frac{3 \bar{m}n_F^S(T)}{2 m_F \tilde{n}_B(T)}} 
\nonumber\\ 
& \times& 
\left[\frac{\pi}{2\epsilon_l} 
\left\{\mbox{sgn}(\epsilon_n+\epsilon_l)-\mbox{sgn}(\epsilon_n)\right\} 
+{\cal O}(\epsilon_l^0)\right]\nonumber\\ 
&\simeq& 
-\frac1{2\epsilon_l \tau(T)} 
\left\{\mbox{sgn}(\epsilon_n+\epsilon_l)-\mbox{sgn}(\epsilon_n)\right\} 
\label{gammab} 
\end{eqnarray} 
where 
\begin{eqnarray} 
\frac{1}{\tau(T)} 
&\equiv& \frac{3\pi\bar{m}}{2m_B}\frac1\beta 
 \Biggl[1+\frac{3\bar{m}n_F^S(T)}{2 m_F \tilde{n}_B(T)}
\nonumber\\ 
& &\hspace{1.5cm} 
-\sqrt{\Bigl(1+\frac{3\bar{m}n_F^S(T)}
{2 m_F \tilde{n}_B(T)}\Bigr)^2-1}\
\Biggr]. 
\end{eqnarray} 
 To calculate $\tilde{\Pi}_{B\;ij}(0, \epsilon_l)$ 
we insert eq.(\ref{GammaB})  
into eq.(\ref{pib}) and do the $q$-integral and $n$-sum  
as in eq.(\ref{gamma}) to get 
\begin{eqnarray} 
\tilde{\Pi}_{B\;ij}(0,\epsilon_l) &\simeq& \delta_{ij}\frac{n_B}{m_B} 
\frac{i\epsilon_l}{ i\tilde{C}(T)\epsilon_l  
+i\tau^{-1}(T)\mbox{sgn}(\epsilon_l)}, 
\end{eqnarray} 
where $\lim_{\epsilon_l \rightarrow 0}\tilde{C}(T)$ is finite. 
 After analytic continuation $\epsilon_l > 0 \to -i\epsilon$ 
and using eq.(\ref{sfd}), 
we finally obtain the resistivity as 
\begin{eqnarray} 
\rho &\simeq& \frac{m_B}{e^2n_B}\frac{1}{ \tau(T)} 
\nonumber\\ &\propto&   
T \Biggl[ 1- \sqrt{\frac{3f_B(|\mu_B|)\bar{m}(1-\delta)}
{2 m_F \delta}} 
 \cdot \frac{\pi(1-\delta)\lambda}{2k_BT}\  \Biggr]. 
\nonumber\\ 
\label{rho} 
\end{eqnarray} 
For $T_{\rm SG} < T < T_{\rm CSS}$, $\lambda = 0$ and 
this result reproduces  the $T$-linear behavior of 
Ref.\cite{ln,iw}. 
For $T $ near and below $T_{\rm SG}$,  
one expects the following behavior; 
\begin{eqnarray}
\lambda &\propto& (T_{\rm SG} - T)^{d},\nonumber\\  
\rho &\propto& T \Bigl[1-c(T_{\rm SG}-T)^{d} \Bigr],
\end{eqnarray}
where $d(\delta,T)$ is a ``critical exponent" in 2D,
which depends both 
on $T$ and $\delta$.   
The downward deviation of $\rho$  
from the $T$-linear behavior is reduced  
with increasing the doping $\delta$. 
In Fig.\ref{rhod}, we plot $\rho$ of eq.(\ref{rho}) with 
various values of $d$. The MFT value $d  = 1/2$  of (\ref{MFT})
is not consistent with the experiment which gives rise to
smooth deviation from the $T$-linear behavior (no discontinuity
in $d\rho/dT$ at $T= T_{\rm SG}$). 
To achieve such a
behavior, one needs $d > 1$.   
This implies that the fluctuation effect of phases of  
$\lambda_{xi}$ beyond the MFT is important to obtain a  
realistic curve of $\rho$.  
 
The data of Ref.\cite{uchida} show that one may fit $\rho$  
in a form $C_0 + C_1 T$ 
for $  T_{\rm SG} < T$. This implies spinon contribution to $\rho$,  
calculated in Refs.\cite{ln,oim}  
as  $\sigma_F^{-1} \sim (k_BT/\mu_F)^{4/3}$,  
is negligibly small compared with  $\sigma_B^{-1} \sim m_Bk_BT/n_B$  
due to higher 
power in $T$ and a small coefficient.  
$\sigma_F^{-1} = 0 $ for $T < T_{\rm SG}$ as explained,  
but the discontinuity at $T_{\rm SG}$ in $\sigma_F^{-1}$ 
is not observable due to its smallness.   
 
The constant part $C_0$, surviving below $T_{\rm SG}$, 
may be attributed to  
scatterings of charged holons with impurities.  
They may be described by $H_{\rm imp} = \sum V_x b^{\dagger}_x 
b_x $, where $V_x$ is a random potential.  
Actually, standard calculations show that it generates   
$T$-independent contribution to $\rho$,   
$\Delta \sigma_B^{-1} \propto 1/n_B  $ at 
intermediate $T$'s.\cite{c0} 
 
 
\subsection{Spin-Gap Effect on Resistivity } 
 
Let us estimate the momentum cutoff $q_c$ of the gauge boson. 
Since the gauge boson
is viewed as a composite particle made of two spinons and of two
holons, it is natural to estimate it by using the cutoff $k_F$
of spinons as
\begin{eqnarray}
q_c &=&2 k_F
\label{qc}
\end{eqnarray}
for small $\delta$ where the spinons dominate over holons.
The qualitative behavior of $\rho$ near $T_{\rm SG}$
is not sensitive to the choice of $q_c$. In fact, 
the exponent $d(T_{\rm SG})$ is independent of $q_c$.

Next, let us consider  the renormalization effect of 
the hopping parameter $t$. 
We assume that the three-dimensional system exhibits   
Bose condensation at the temperature scale of  
\begin{eqnarray}
T_{B} &\simeq&  2\pi \frac{n_B}{m_B} = 4\pi t \chi \delta,
\end{eqnarray}

$T_{B}$ as the observed   
$T_{\rm c}$ in the lightly-doped region. Since $t \sim 0.3$ eV gives  
rise to $T_B \sim 3000$ K at $\delta \sim 0.15$, 
one needs to use an effective hopping parameter $t^* \sim 0.01$ eV  
in place of $t$ so  
as to obtain a realistic $T_{\rm c} \sim 100 $K \cite{tstar}. 
 
\begin{figure} 
  \begin{picture}(200,120) 
    \put(0,0){\epsfxsize 200pt \epsfbox{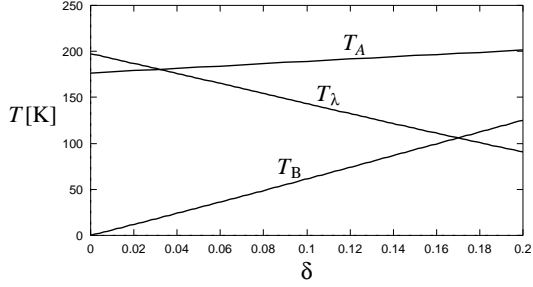}} 
  \end{picture} 
  \caption{ 
    Mean-field phase diagram of the t-J model. 
    $T_{BC}$ is the Bose condensation temperature. 
    $T_\lambda$ is the spin-gap onset temperature.  
    $T_A$ is the root of $T=\Theta(T)$ at which $d(T)$ 
    diverges. We chose 
    $t^*=0.01$ eV, $J = 0.15$ eV, and  
    $\omega_\lambda = \pi J \chi/(2e^{\gamma})$. 
    } 
\label{phase} 
\end{figure}

\begin{figure} 
  \begin{picture}(200,200) 
    \put(0,0){\epsfxsize 170pt \epsfbox{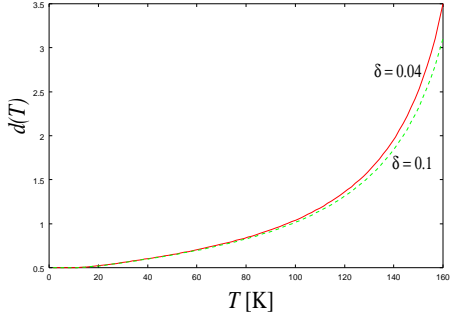}} 
  \end{picture} 
  \caption{ 
Exponent $d(T)$ vs $T$ for $\delta  = 0.04$ and $0.1$.  
} 
\label{dT} 
\end{figure}

In Fig.\ref{phase}, we present the phase diagram calculated by
using these parameters, where 
 $T_{\rm SG}$ is the spin-gap on-set temperature, 
$T_A$ is  the temperature at which the mass exponent $d$ diverges, 
and $T_B$ is the bose condensation temperature. 
In Fig.\ref{dT}, we plot the $T$-dependence of the exponent $d(T)$. 
In Fig.\ref{rho}, we plot  $\rho$ as a function of $T$
for several $\delta$'s. 
As explained, the curves reproduce the experimental data much 
better than those with $d = 1/2$ of the MF result, showing  smooth 
departures from the $T$-linear curves, i.e., $d (T_{\rm SG}) > 1$ 
for the region of interest, $0.05 \lesssim \delta \lesssim 0.15$.

 
\section{Discussion} 
\setcounter{equation}{0} 
In this paper, we studied the effective gauge 
theory and the phase structure of the slave-boson t-J model. 
We reviewed our analytical studies which indicates the CSS at  
low temperatures as supported by experiments. 
We also explained that the spin-gap state corresponds 
to the Higgs phase in the effective gauge theory. 
Both analytical and numerical studies show that there is a phase 
transition to the Higgs phase in $3D$. 
In the second half of this paper, we summerized the calculation 
of the resistivity especially in the spin-gap phase.
The results of $\rho(\delta,T)$ are consistent with the 
experimental observations. This may support that our treatment of 
gauge-field fluctuations in a quasi-2D system
by the variational treatment of compactness 
is suitable to describe the spin-gap state in the t-J model.

Concerning to the dimensionality, we cite the work of 
Su et al.\cite{Su}. They consider the pure 2D SB t-J model,
and apply the dualty transformation for bosonized spinons
as in Ref.\cite{BFN}. They calculate $\rho$ in the
spin-gap state using the disorder parameter, the density of
vortex condensation, $v(x)$, which is nonvanishing 
in the spin-gap state as explained in Table 2.
(For more details, see (\ref{ab}).)
After fitting several parameters
to the experimental data,  they plotted $\rho$.
We stress that their approach is dual to ours,
but fails to locate $T_{\rm SG}$. This makes their parameter
fitting rather obscure.

Finally, we address here on the observability of holons and 
spinons in the CSS state because it often brings some 
misunderstandings.  
In Sect.2B, we explained that the gauge dynamics is in the 
deconfinement phase or the Coulomb phase below $T_{\rm CSS}$. 
Then one may ask if the holon and spinon appear as quasiparticles, 
though physical quantities are all gauge-invariant. 
Deconfinement of the slave particles 
{\it does not} necessarily imply that these particles are 
``observed" each by each. This point is often misunderstood. 
 
To see this,  it is useful to recall QED.  
In QED in $3+1$-dimensions, genuine asymptotic states are 
constructed by the gauge-invariant operators \cite{KF} like 
\begin{eqnarray}
\psi^\dagger(x) \sum_P w(P)\exp(ie\int_P dx_i A^{\rm EM}_i),
\label{electron}
\end{eqnarray} 
where $\psi(x)$  is the electron operator and $A^{\rm EM}_i$ 
is the photon operator. 
$\sum_P$ implies superposition over different paths $P$
starting from $x$ to infinity with a suitable weight $w(P)$.
$w(P)$ characterizes the configuration of electric field
since the exponential factor acts as a creation operator
of electric flux along the path $P$.
In the deconfinement phase like the Coulomb phase, 
the correlation functions of  
the above operator can be expanded in powers of $e$ 
and at each order of $e$ 
the infrared singularities are concelled out with each other. 
As the electrons in QED, the holons and spinons in the CSS state
should accompany 
clouds of the (soft) gauge bosons and genuine quasiparticles 
should be described by gauge-invariant operators like 
(\ref{electron}). However, explicit construction of such operators
is a future problem.\cite{baskaran}  
 
Here we should mention that there is a point that  
distinguishes the quasiparticles  of the t-J model from those of QED. 
Since the t-J model is the model of electrons, any physical 
quantities 
such as conductivity are expressed by the electron operators, and 
therefore gauge-invariant (here the gauge invariance means the 
invariance under a local rotation of the phases of slave particles, 
and nothing to do with the EM gauge symmetry). The deconfinement of 
slave particles means that these physical quantities that are 
expressed by slave particles are (approximately) calculated 
by pertubative calcualtion with respect to the auxiliary 
gauge field which couples to the slave particles, 
becuase fluctuations of the gauge field is not large. 
In this sense, the holons, the spinons and the gauge bosons can be 
regarded as (weakly interacting) quasiexcitations at low energies 
that live only {\it inside} of the material.  This point is 
different from the well known gauge theories in particle physics 
such as QED, in which one can prepare quasiparticles at low 
energies as observable incoming and outgoing particles at 
infinitely separated positions in space. Thus, we think that 
one needs further studies to propose a clever method to directly 
measure the charges of the slave particles. 
 
Associated with the above problem of observability of the slave 
partices, the following question \cite{nayakreply} may arise; 
Since the electron 
operator is constructed as Eq.(\ref{slaveboson}) 
one may assign the EM charges of holons and 
spinons freely as long as the electron has the 
known EM charge $e (<0)$. 
Because deconfinement implies that the EM charges of holons and 
spinons can be observed separately, how does this freedom be fixed ? 
If this freedom survives, it causes nonsensical result like 
fractional charges of holons and spinons.   A similar question 
has been addressed in Sect.5A for the conductivity, and the 
answer there was that the conductivity is independent of this freedom. 

Let us consider this question in detail here. 
This question makes sense if the  
EM charges, $Q_h,\ Q_s$, which one may assign to slave particles,  
were the absolute charges that are measured from the vacuum (the  
state with no electrons). However,  the ambiguity in  
assigning the charges of slave particles just reflects our freedom  
to select the reference state from which their charges are measured.  
For example, if we measure the charges of a holon, $Q_h$, and a  
spinon, $Q_s$, from the half-filled state, we have  
$Q_h = -e, Q_s = 0$,  
while if they are measured from the vacuum, then $Q_h = 0, Q_s = e$.   
More generally, if they are measured from the state of average 
absolute charge $-c\; e$ per site ($c$ is a real number), 
one has $Q_h = 0 - (-c\; e) =
c\; e, Q_s = e - (-c\; e) =  (1+c)\; e$. 
These expressions are nothing
but what we explained at (\ref{shift1}) and (\ref{shift2}) in Sect.5A. 
Thus they are physically equivalent. 
It is also instructive to see  
that the electron charge per site,  
\begin{eqnarray} 
Q_C &= &e \sum_{\sigma}C^{\dag}_{i\sigma} C_{i\sigma}    
= e \sum_{\sigma}f^{\dag}_{i\sigma} f_{i\sigma} =  
e ( 1 - b^{\dag}_{i} b_{i}), 
\label{charge2} 
\end{eqnarray}  
depends on the state itself. 
For the half-filled state,  $\prod_{i} f^{\dag}_{i} |0\rangle$  
($|0\rangle$ is the state with no slave particles),  we have 
$ Q_C = e$,  while for the vacuum state, $\prod_{i} b^{\dag}_{i} 
|0\rangle, Q_C = 0$. 
The EM charge is an additive quantity, and what we actually measure
as ``charges" in the experiments are always  the differences of the 
charges. 
For example, by applying the  
electron variable, $ C_{i\sigma} =  b_i^{\dag} f_{i\sigma}$, to a  
state, the EM charge changes by $Q_h - Q_s = c e - (1+c) e = -e$.  
This difference is of course independent of 
the reference state itself and has a physical meaning.

 
\begin{center} 
{\bf Acknowledgments} 
\end{center} 
We appreciate the discussion with K. Sakakibara 
on the CSS and the problem of observation of slave particles. 
One of the authors(I.I.) acknowledges K.I.Kondo for helpful 
discussions.  
 

\renewcommand{\theequation}{\Alph{section}.\arabic{equation}} 
\appendix  
 
\setcounter{equation}{0}

\section{Phase Structure of the Compact Abelian Higgs Model} 
In Sect.4, we derived the dual-variable represenation
(\ref{Zdual}) of the noncompact Abelian Higgs model (\ref{ZAH}). 
Similar analysis is possible in the continuum field theory of 
the {\it compact} Abelian Higgs model (\ref{ZAHC}).
Nagaosa and Lee\cite{NLdual} studied this model
in 3D continuum and concluded that the system is always in
the confinement phase. 
In this Appendix, we point out that their
argument is insufficient and the phase transition into the
deconfining-Higgs phase (i.e., the superconducting phase
with a massive gauge field in their context) certainly exists. 

The action is given by 
\begin{equation} 
S=\int d^3x\Big[{\rho\over 2}\Bigl
(\vec{\nabla} \phi (x)-\vec{A}(x)\Bigr)^2 
+{\kappa\over 2}(\vec{\nabla} \times \vec{A})^2\Big], 
\label{actioncont} 
\end{equation} 
where we consider the $3D$ continuum system, 
and $\phi$ is the phase of the Higgs field  and $\vec{A}$ is
the vector potential. 
Duality transformation for (\ref{actioncont}) can be performed as 
in the lattice model. 
We take the U(1) gauge group as compact and allow 
singular configurations of $\vec{A}$ as
\begin{eqnarray}  
\vec{\nabla}\cdot\vec{\nabla} \times \vec{A}=4\pi Q(x),
\label{Q}
\end{eqnarray}
where $Q(x)$ is the instanton (monopole) 
density as  (\ref{instantonsource}) in the lattice model. 
The phase $\phi(x)$ of the Higgs  field is also compact, 
so singular 
configurations with nonvanishing vorticity are allowed;
\begin{eqnarray}  
\vec{\nabla} \times \vec{\nabla} \phi=2\pi \vec{J}.
\end{eqnarray}    
$\vec{J}$ is the vortex current as (\ref{vortex}) 
in the lattice model. 
If the radial degrees of freedom of the Higgs field are introduced,
vortices can appear as regular solitons of the field equations, 
which are called the Nielsen-Olesen vortex\cite{NO}.
In any way,
it is standard to introduce a complex scalar ``vortex field" 
$\psi_V$, whose world lines are identified with $\vec{J}$.
 The field $\psi_V(x)$ is viewed as 
the creation(annihilation) operator of a vortex.
Then $\psi^\dagger_V(x) \psi_V(x)$ measures the vortex density.  
$\psi_V$ interacts with each other via a short-range interaction 
as in (\ref{Zdual}). 

One can sum up instantons by the dilute gas approximation.
To do this, one introduces a phase field $\varphi(x)$. In fact,
the relation (\ref{Q}) is respected by the Lagrange
multiplier field $\varphi(x)$ as
\begin{eqnarray}
&&\int_{-\pi}^{\pi} d\varphi(x)
\exp\Bigl(i\varphi(x)
(\frac{1}{4\pi}\vec{\nabla}\cdot\vec{\nabla} \times \vec{A}(x)
- Q(x))\Bigr),
\label{AQ}
\end{eqnarray}
and the summation over the instanton $Q(x) =0, \pm1$ gives rise to
a factor,
\begin{eqnarray}
&&\sum_{Q(x) = 0, \pm1} e^{-i\varphi(x)Q(x)}\ z^{|Q(x)|}
[\psi_V(x)]^{Q(x)}
\nonumber\\
&=& 1+ z \Bigl(e^{-i\varphi(x)}\psi_V(x) 
+ e^{i\varphi(x)}\psi^\dagger_V(x)\Bigr)
\nonumber\\
&\simeq& \exp\Bigl( z (e^{-i\varphi(x)}\psi_V(x) + 
e^{i\varphi(x)}\psi^\dagger_V(x)) \Bigr),
\label{varphipsi}
\end{eqnarray}
where $z$ is the fugacity of instantons, and 
the first line describes that each instanton $Q(x)$ is
connected with the vortex current generated by $\psi_V(x)$.
($\psi_V(x)^{Q(x)}  \equiv (\psi_V(x)^{\dagger})^{-Q(x)}$
for $Q(x) < 0$.) 
As the factor $\exp(i\varphi Q)$ indicates, the angle
$\varphi(x)$ is conjugate to the integer $Q(x)$. 
Thus, there holds the uncertainity principle,
\begin{eqnarray}
\Delta \varphi(x) \Delta Q(x) \ge 1.
\end{eqnarray}
This implies that the following two phases are possible;
\begin{eqnarray}
{\rm (i)\ \ \ \  Confinement:}\ 
& \Delta \varphi(x) = 0; & \Delta Q(x) = \infty, \nonumber\\
{\rm (ii)\  Deconfinement:}\ & \Delta \varphi(x) = \infty;\ 
&  \Delta Q(x) = 0.
\label{iii}
\end{eqnarray}
(i) is the confinement phase where the instanton condenses,
hence $A_i$ fluctuates violently $\Delta A_i = \infty$  
as seen by (\ref{Q}).
(ii) is the deconfinement phase where no instantons appear 
$Q(x) =0, \Delta Q(x) = 0$, so $\Delta A_i = 0$.

Next, we note that the Higgs field
$\exp(i \phi)$ and the vortex field  $\psi_V$ are ``conjugate"
each other in the sense that the following two cases are possible;
\begin{eqnarray}
{\rm (a)\ Disorder-Vortex:}\ & 
\langle \psi_V \rangle \neq 0;\ & 
\langle \exp(i \phi) \rangle = 0,\nonumber\\
{\rm (b)\ \ \ \ \ \  Order-Higgs:}\ & \langle \psi_V \rangle  = 0;\ &
\langle \exp(i \phi) \rangle \neq 0. 
\label{ab}
\end{eqnarray}
Actually, when vortices proliferate with finite density
$\langle \psi^\dagger_V(x) \psi_V(x) \rangle$, 
$\langle \psi_V(x) \rangle \neq 0$, while
$\Delta \phi = \infty$ and so $\langle \exp(i \phi) \rangle = 0$.
This is the case (a), which one may call the disorder  phase, 
$\Delta \phi = \infty$, or the vortex phase.
 On the other hand, when vortices are rare, $\langle \psi_V(x) 
 \rangle = 0$, while $\langle \exp(i \phi) \rangle \neq 0$ due to 
 $\Delta \phi = 0$. This is the case (b), which on may call
 the  ordered phase, $\Delta \phi = 0$, or the Higgs phase.
This Higgs-vortex duality is summerized in Table 2, 
where the vortx density
$v(x) = \langle \psi^{\dagger}_V(x) \psi_V(x) \rangle$.

The continuum action can be expressed 
in terms of $\varphi$ and the dual variables as follows;\cite{NLdual} 
\begin{eqnarray} 
S &=&\int d^3x \Big[{1\over 2\rho}(\vec{\nabla}\times \vec{c})^2+ 
{1\over 2\kappa}(\vec{c})^2+{1\over 2\kappa}(\vec{\nabla} \varphi)^2 
\nonumber \\ 
&+&{1\over 2}\psi_V^\dagger[-K(\vec{\nabla}+i\vec{c})^2+M^2]\psi_V 
+g(\psi^\dagger_V\psi_V)^2  \nonumber \\ 
&-&z (\psi^\dagger_Ve^{i\varphi}+\psi_Ve^{-i\varphi}) 
\Big]. 
\label{Zdualcont} 
\end{eqnarray}
By integrating over $\psi_V$, its world lines (the vortex current)
are generated, and with each line segment the current $\vec{c}$ 
is coupled.
Integration over the force-mediating field $\vec{c}$ 
produces interactions among the vortex currents.
The mass of the vortex field is given as 
$M^2=4\pi^2\rho D(0;m^2)-\ln \mu$ as in the lattice model. 
There is also a short-range repulsive interaction 
($g > 0$) between vortices. 
 
To study the phase srucutre, Nagaosa and Lee\cite{NLdual} 
introduced a new field $\psi'_V=\psi_Ve^{-i\varphi}$ 
by a field redefinition or a ``gauge transformation". 
The last line of the action (\ref{Zdualcont}) then generates
a linear term of $\psi'_V$. Due to this term,
they simply concluded that $\psi'_V$ always condenses  
\begin{eqnarray}
\langle \psi'_V \rangle \neq 0
\end{eqnarray}
regardless of the value of $M^2$. From this, they argued that   
the system is always in the vortex condensed phase of (a)
of (\ref{ab}), and
there occurs no phase transition into the Higgs phase (b) of
 (\ref{ab}). 
 
However, their argument above using the field redefinition
or gauge transformation is not correct.  
A good counter example is the gauge-Higgs model itself. 
In the original variables, the gauge-Higgs coupling is given by 
\begin{equation} 
e^{i\phi(x+i)}e^{iA_i(x)}e^{-i\phi(x)}+\mbox{H.c.} 
\label{gaugeHiggs} 
\end{equation} 
We can fix the gauge to the unitary gauge, i.e.,
$\exp(i\phi(x)) = 1$ and then the linear term  of the 
exponentiated gauge field $\exp(iA_i(x))$ appears in the action,
i.e., $S\sim UUUU+U$. 
But this linear term does not necessarily lead to 
$\langle\exp(iA_i(x))\rangle \neq 0$ which is the condition
that the system is in a nonconfining phase, like the Coulomb phase
or the Higgs phase. 
For example, a MFT of this model with a variational action of
single-link form $S \sim U$ offers us two 
possibilities,\cite{Elitzur}
(i) $\langle U \rangle \neq 0$ for the Coulomb or the Higgs phase
or (ii) $\langle U \rangle = 0$ for the confinement phase,
and predicts a phase transition between (i) and (ii).

To confirm this point, let us analyse the system 
(\ref{Zdualcont}) in a more straightforward manner
referring to Polyakov's result.\cite{polyakov}
He considered the dynamics of the pure compact U(1) gauge model 
in $3D[(2+1)D]$, and showed that the system is always in 
the confinement phase. After integratong over the gauge field
(instantons), the action $S_{U(1)}$ is given by
\begin{equation} 
S_{U(1)}=\int d^3x \Big[{1\over 2\kappa}(\vec{\nabla}\varphi)^2 
-z\cos \varphi\Big],
\label{SU1} 
\end{equation} 
where the field $\varphi$ here again conjugates to instanton density
and mediates the force among instantons.
One can approximate the relevant potential term $\cos \varphi$ as  
$\cos \varphi \sim \varphi^2$ to find that the vacuum is 
given by $\varphi=0$ (mod $2\pi$). 
This obviously means $\Delta \varphi \ll 1$, and so
a condensation of instantons; 
the state (i) of (\ref{iii}) above. 
Thus the system is in the confinement phase. This is the result
of Polyalov.\cite{polyakov} 

Now let us return to the present model (\ref{Zdualcont}) and
consider two cases,  $M^2 < 0$ and  $M^2 > 0$, separately.
We assume $z$ is sufficiently small, so that the 
renormalization effect of $M^2$ is negligible, although
the last $z$ term gives rise to a negative renormalization to $M^2$.

For $M^2 < 0$, the vortex field condenses  
$\langle \psi_V \rangle=\psi_{0} \neq 0$ 
because the effective action of $\psi_V$ has a negative curvature at
the origin as in the standard Ginzburg-Landau theory with the 
{\em global} U(1) symmetry. 
Then the field $\varphi$ acquires a potential term like  
$z\psi_{0}\cos \varphi$ where we assumed that $\psi_{0}$ is real 
without loss of generality. The instanton action $S_\varphi$ 
reduces to that of the pure U(1) gauge 
model (\ref{SU1}), and the Polyakov's result applies;
condensation of instantons occurs and the confining phase is 
realized.\cite{NGboson}  This phase corresponds to 
the confinement-vortex phase (ia) of 
(\ref{iii}) and (\ref{ab}).

For $M^2 > 0$ we can safely integrate out the vortex 
field because there are only short-range interactions between 
vortices. 
Then the resultant action describes the dynamics of instantons 
or of their conjugate variable $\varphi(x)$. 
The action of $\varphi(x)$ is given as  
\begin{eqnarray} 
S_\varphi &=& {1\over 2\kappa} \int d^3x  (\nabla\varphi)^2 
-z^2\int d^3xd^3y e^{i\varphi(x)}e^{-M|x-y|}e^{i\varphi(y)}. 
\nonumber\\
\label{Sinst} 
\end{eqnarray} 
The second term in (\ref{Sinst}) originates from the fact that 
an instanton and an anti-instanton is connected by a vortex flux 
tube via (\ref{instantonsource}). 
We recognize that, for large $M$, the second term of (\ref{Sinst})
is approximated by a derivative term, hence  
\begin{eqnarray} 
S_\varphi &\to& {1\over 2\kappa} \int d^3x  (\nabla\varphi)^2 
+z^2\int d^3x (\nabla\varphi)^2. 
\nonumber\\
\label{Sinst2} 
\end{eqnarray} 
There is no potential term of $\varphi$ in 
(\ref{Sinst2}) and $\varphi$ does not have a definite value,
but fluctuates randomly (for sufficiently large $\kappa$ 
and small $z^2$). 
This is the state (ii) above, leading to a deconfinement phase.  
Here, we have $\langle \psi_V\rangle =0$  
because the large fluctuation of $\varphi$ makes the last 
linear term in  $\phi_V(x)$ irrelevant;
$\langle \exp(i\varphi) \rangle =0$.
From (\ref{ab}), this means that the original Higgs 
field has a coherent phase. This phase corresponds to 
the deconfinement-Higgs phase (iib) of 
(\ref{iii}) and (\ref{ab}).

The above arguments indicate that there is
a phase transition from the confining to the Higgs phases. 
This is actually found now in the numerical 
studies.\cite{Higgs} 
On the other hand, for the compact gauge-Higgs  
model with fixed radius of the Higgs field, Fradkin and Shenker 
showed\cite{FS} 
that the confinement and Higgs phases are connected if the Higgs 
charge is fundamental $q=\pm1$. 
The above two observations can be reconciled if the critical line 
terminates at some point in the $(\kappa-\rho)$ plane.
If the radius of the Higgs field is also a dynamical variable, 
the critical line does not terminate as the parameter space is now 
three dimensional instead of two.


\eject 

\begin{tabular}{|c|c|c|}\hline
Phase & $T$ & $V(r)$ \\ \hline
Confinement & $T_{\rm CSS} < T $ & $\propto r$ \\ \hline
Coulomb & $T_{\rm SG} < T < T_{\rm CSS}$ & $\propto \frac{1}{r}$ 
\\ \hline
Higgs & $T < T_{\rm SG}$ & $\propto \exp(-m_A\; r)/r$ \\ \hline 
\end{tabular}\\

Table 1. Phases of the gauge dynamics of the SB t-J model.
$V(r)$ is the potential energy between two external charges
separated by a distance $r$.  

\vspace{1cm}

\begin{tabular}{|c|c|c|c|}\hline
Phase & $T$ & $\lambda(x)$ & $v(x)$ \\ \hline
Coulomb & $T_{\rm SG} < T < T_{\rm CSS}$ & $0$ & $\neq 0$ \\ \hline
Higgs & $T < T_{\rm SG}$ & $\neq 0$ & $0$ 
\\ \hline 
\end{tabular}\\

Table 2. Comparison of the ordinary represenation and 
the dual representation of the 3D (quasi-2D) 
noncompact Abelian Higgs model
for the transition at $T_{\rm SG}$ into the spin-gap state.
The order parameter of the ordinary representation is
the spin-gap amplitude $\lambda(x)$, while the disorder parameter
in the dual representation is the vortex density $v(x)$.
In the pure 2D system, $T_{\rm SG} = 0$ and only the Coulomb phase 
exists, in which $v(x)$ is always nonvanishing.
  

\vspace{1cm}
 
\vspace{-3cm} 
 
\begin{figure} 
  \begin{picture}(200,200) 
    \put(0,0){\epsfxsize 200pt \epsfbox{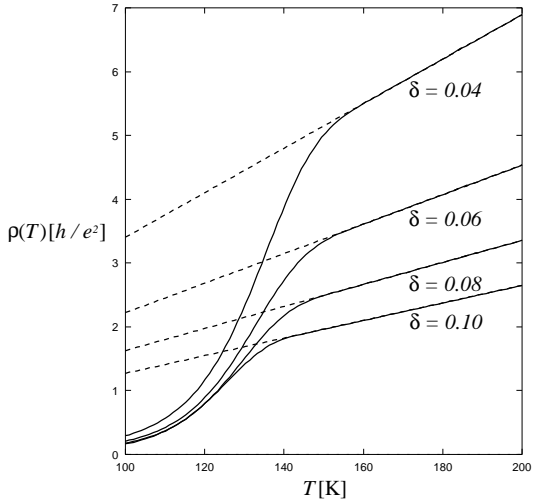}} 
  \end{picture} 
  \caption{ 
    Resistivity $\rho(\delta,T)$ in $h/e^2$ for several $\delta$'s    
with the parameters chosen in Fig.2.  
The dotted lines represent the case of $X(T)  = 0$ in (1). 
The exponent $d(T_{\rm SG})$ decreases 
as 16.4, 4.8,  2.8,  2.0, 
as $\delta$ increases.  
} 
\label{rho} 
\end{figure} 
 

\begin{references}
\bibitem{ourexp}
We wrote this explanation in a letter to the editor.
The Referee recommended it to publish, but it was 
not included in the 
comment\cite{nayakcomment}-reply\cite{nayakreply}.
\bibitem{3dcoupling} 
For a more realistic Hamiltonian with anisotropic couplings
reflecting quasi-two-dimensionality, see
(\ref{quasi2D}). We present the isotropic expressions below
because most of the argument does not require the explicit 
anisotropic formula.
\bibitem{chiestimate} 
The first term of Eq.(\ref{chiestimate}) is obtained by 
modifying the form of 
Fermi curve by the square of volume $\pi^2(1-\delta)a^{-2}$.
This is a good approximation with errors below 1\% up to $\delta = 0.3$
\bibitem{supercharge}
Because the carriers
of superflow are NN hole pairs, the electromagnetic charge of 
each carrier $Q_{\rm super}$ should be $-2e\ (> 0)$. 
However, the present SB MFT ignores the pairing of two holons, 
predicting $Q_{\rm super} = -e$ incorrectly. On the other hand,
the SF approach \cite{IM} describes holon pairings and predicts 
$Q_{\rm super} = -2e$ correctly.
\bibitem{asymmetricxy}
The asymmetric XY spin model for XY spins 
$(S^1_x, S^2_x) \equiv (\cos \theta_x, \sin \theta_x)$ 
has the action
$\sum_{x,\i} [c_1 S^1_x S^1_{x+i} + c_2 S^2_x S^2_{x+i}]$,
where the asymmetry is caused by the spin gap and
implies $c_1 \neq c_2$. If no spin gap exists, the model reduces
to the usual (symmetric) XY model (with $c_1 = c_2$).
\bibitem{NGboson}There appear massless Nambu-Goldstone bosons
as a result of the spontaneous breaking of the global U(1) symmetry.
Then one may wonder about the stability of the ``ground state" with
$\varphi=0$. However it can be verified that the massless boson
does not contribute to the mass of  $\varphi$ at $O(z^2)$.
\end{references}
\end{document}